\documentclass[preprint,12pt]{elsarticle}

\usepackage{epsfig}

\usepackage{amssymb}
\usepackage{amsthm}
\usepackage{xcolor}
\usepackage{lineno}
\usepackage{caption}
\usepackage{subcaption}

\usepackage[utf8]{inputenc}
\usepackage{graphicx}
\usepackage{setspace}
\usepackage{natbib}
\usepackage{float}
\usepackage{url} \Urlmuskip=0mu plus 1mu 
\usepackage{tabto}
\usepackage{amssymb,amsmath,amsthm, mathtools}
\usepackage{amssymb,amsmath,amsthm, mathtools}
\usepackage{graphicx,url}
\usepackage{physics}
\usepackage{graphicx,wrapfig,lipsum}
\usepackage{amsmath}
\usepackage{amsfonts}
\usepackage{amssymb}
\usepackage{epstopdf}
\journal{???}
\renewcommand{\vec}[1]{\mathbf{#1}}
\newcommand{\mat}[1]{\mathbf{#1}}
\begin{document}

\begin{frontmatter}



\title{Complete modeling of hydrodynamic bearings with a boundary parameterization approach}



\author[inst1]{J.A. Mota}

\affiliation[inst1]{organization={Institute of Computing},
            addressline={UFRJ}, 
            city={Rio de Janeiro},
            country={Brazil}}

\author[inst2]{D.J.G. Maldonado}
\author[inst1]{J.V. Val\'erio}
\author[inst2]{T.G. Ritto}

\affiliation[inst2]{organization={Department of Mechanical Engineering},
            addressline={UFRJ}, 
            city={Rio de Janeiro},
            country={Brazil}}

\begin{abstract}

The present work aims to revisit the simplifications made in the Navier-Stokes equations for the flow between two cylinders with a small thickness of lubricating oil film. Through a dimensionless analysis, the terms of these equations are mapped and ordered by importance for the hydrodynamic bearing application. An effective parameterization of the geometry is proposed, enabling a more detailed description of the problem and its adaptation to other contexts. At the end, an elliptical partial differential equation is reached and solved by the centered finite difference method, whose solution is the pressure field between the cylinders. To illustrate the effectiveness of the proposed approach, the model is applied to hydrodynamic bearings, where the pressure field and some parameters resulting from it, such as stiffness and damping coefficients, are computed. Based on the facilities offered by the parameterization of the geometry, two different configurations are presented: (1) elliptical and (2) worn bearings. Their responses are evaluated and a comparative analysis is performed. The modeling exposed in this text, as well as all its simulations were developed to integrate \textit{Ross-Rotordynamics}, an open library in \textit{Python}, available on the \textit{GitHub} platform.

\end{abstract}



\begin{keyword}
Lubrication Theory \sep Geometry parameterization \sep Numerical Simulation \sep Hydrodynamic Bearings \sep Dynamic Coefficients
\end{keyword}

\end{frontmatter}


\section{Introduction}
\label{sec:Introduction}

The dynamics of rotating machinery has been actively studied since the beginning of the last century. Many industries rely on the  performance of such machines. Any failure or malfunction affects the productivity of the process. Thus, it is of great importance to know in advance the behavior of the machine under different operating conditions. In that study, special attention has been drawn to elements with fluid-induced forces. This is the case of hydrodynamic bearings and annular seals. These components are composed of a stator, a rotor and a fluid between them that reacts to external loads applied to the rotor. Depending on the geometry and the operating conditions, those forces may induce high vibration on the rotor. In general, the prediction of the rotor vibration under a certain operating condition is the main concern of the investigations.

Journal bearings play an important role in rotating machines since they support the rotor and the external loads applied to it. Since the reaction forces of the fluid may increase the vibration of the machine, it is necessary to understand the bearings forces in order to know the behavior of the machine under different conditions. Based on this information, predictions about malfunctions or different observed phenomena can be understand and corrected in advance.

One of the main concerns of the rotordynamics are the high vibrations due to instabilities and resonances. Rankine \cite{rankine_centrifugal_1869}, Dunkerley \cite{dunkerley_whirling_1893}, Föppl \cite{foppl_problem_1895} and Jeffcott \cite{jeffcott_lateral_1919} were the pioneers that developed investigations on the critical speeds of a spinning rotor supported by bearings. The objective of their studies was the stability of the rotor's motion beyond the first critical speed. Since then, stability has been one of the most important characteristics of rotating machines; specially today, where there is a need of high performance machines with reliable operating conditions. A more detailed review of bibliography can be found in the books of Childs \cite{childs_turbomachinery_1993}, Muszynska \cite{muszynska_rotordynamics_2005}, Tiwari \cite{tiwari_rotor_2017}, Vance \cite{vance_rotordynamics_1988} and Ishida and Yamamoto \cite{ishida_linear_2013}.


Regarding hydrodynamic bearings, the research began in the 1880s, with the experimental investigations of Beauchamp Tower (1845-1904) and Nicolai P. Petrov (1836-1920). However, the theoretical basis for their experimental observation was only presented by Osborne Reynolds (1842- 1912). The three of them can be considered the founding fathers of the hydrodynamic lubrication and their findings are still used these days. Sommerfeld \cite{sommerfelda:1904} found an analytical expression for infinitely long bearings, but only Swift \cite{swifth.w.1932} and Stieber \cite{stieberw.1933} used realistic boundary conditions to the Reynolds equation. Ocvirk \cite{ocvirkf.w1952} developed a detailed solution to infinitely short bearings, which is still widely used today. For the finite bearing, a strong effort was given by Cameron and Wood \cite{camerona.woodl1949} with numerical calculations without the aid of electronic devices. After the arrival of computers, Pinkus \cite{pinkuso1956} was the first researcher to use the computer for the numerical calculations of the solutions from circular bearings, elliptical bearings and a three lobes bearing. In 1981, Singhal \cite{singhalg.c.1981} applied finite differences to the Reynolds equation.


In addition to the lubrication theory, the studies carried out by Pina and Carvalho \cite{Pina:05}, Andrade \cite{selma:99} and Queiroz \cite{queiroz:2018} are relevant because their work has inspired the parameterization used in the description of the geometry. The well-known collection of lecture notes recorded by San Andres \cite{SanAndres:note4}, especially the $ 1 $ to $ 5 $ notes, were essential for understanding the context of hydrodynamic bearings. The works of Frene et al \cite{Frene}, Hamrock \cite{Hamrock} and Ishida and Yamamoto \cite{Yamamoto} stand out, for this text, in what concerns the approximations of the Reynolds equation. The three authors present basic concepts of hydrodynamic bearing geometry and use the approximate equation for short and infinitely long bearings. Finally, the research of Machado and Cavalca \cite{machado:2015} presents solutions for bearings of different geometries. Machado \cite{machado:2011} analyzes dynamic and operational characteristics of different configurations of hydrodynamic radial bearings: a cylindrical, an elliptical and a trilobular. Already Machado and Cavalca \cite{machado:2015} expose a numerical model that characterizes the bearing with wear on its wall, analyzing its influence on the dynamic response of the rotor.

In this article, a study of a fluid film flow between two surfaces is carried out in a novel way. A new and effective parameterization of the geometry is presented to describe the journal bearing, allowing a more detailed description of the problem. Instead of describing the geometry in terms of a gap function, this work describes the geometry of both the rotor and stator. Thus, different type of bearings can be modeled by simply defining the desired shape for each surface. Once the domain is defined, the pressure distribution of the fluid is obtained by means of the lubrication theory. First, the governing equations of a differential fluid element are determined by the Navier-Stokes equations and they are simplified to obtained the Reynolds equation, whose solution is the pressure field of the fluid flow. This equation is discretized by means of the finite differences method. The pressure field is then obtained and verified with results available in the literature. Afterwards, the fluid forces are linearized around the equilibrium point of the rotor and the dynamic coefficients (stiffness and damping) are computed by the least squares method. This methodology is tested for three different geometries: (1) a cylindrical bearing, (2) a lemon bearing and a (c) bearing with a worn stator. More details about the modeling presented in this text can be found in Mota \cite{Mota:20}.

This article is organized as follows. In Section 2, the description of the fluid flow problem is stated and detailed. In Section 3, the model of the fluid flow is described and the governing equation relating the pressure field and the geometrical characteristics is  obtained. In Section 4, the pressure field is obtained by using the finite differences method. Then, in Section 5 the reaction forces of the fluid film are computed by a numerical integration of the pressure field in the whole domain. Section 6 shows the stiffness and damping coefficients obtained with the least-square method. In Section 7, the model is explored and three geometries are analyzed: a short bearing, an elliptical bearing and a worn bearing. Finally, in Section 8, the conclusions are thrown.


\section{Problem description}
\label{sec:geometria}

\begin{figure}
    \centering
    \includegraphics[width=.5\textwidth]{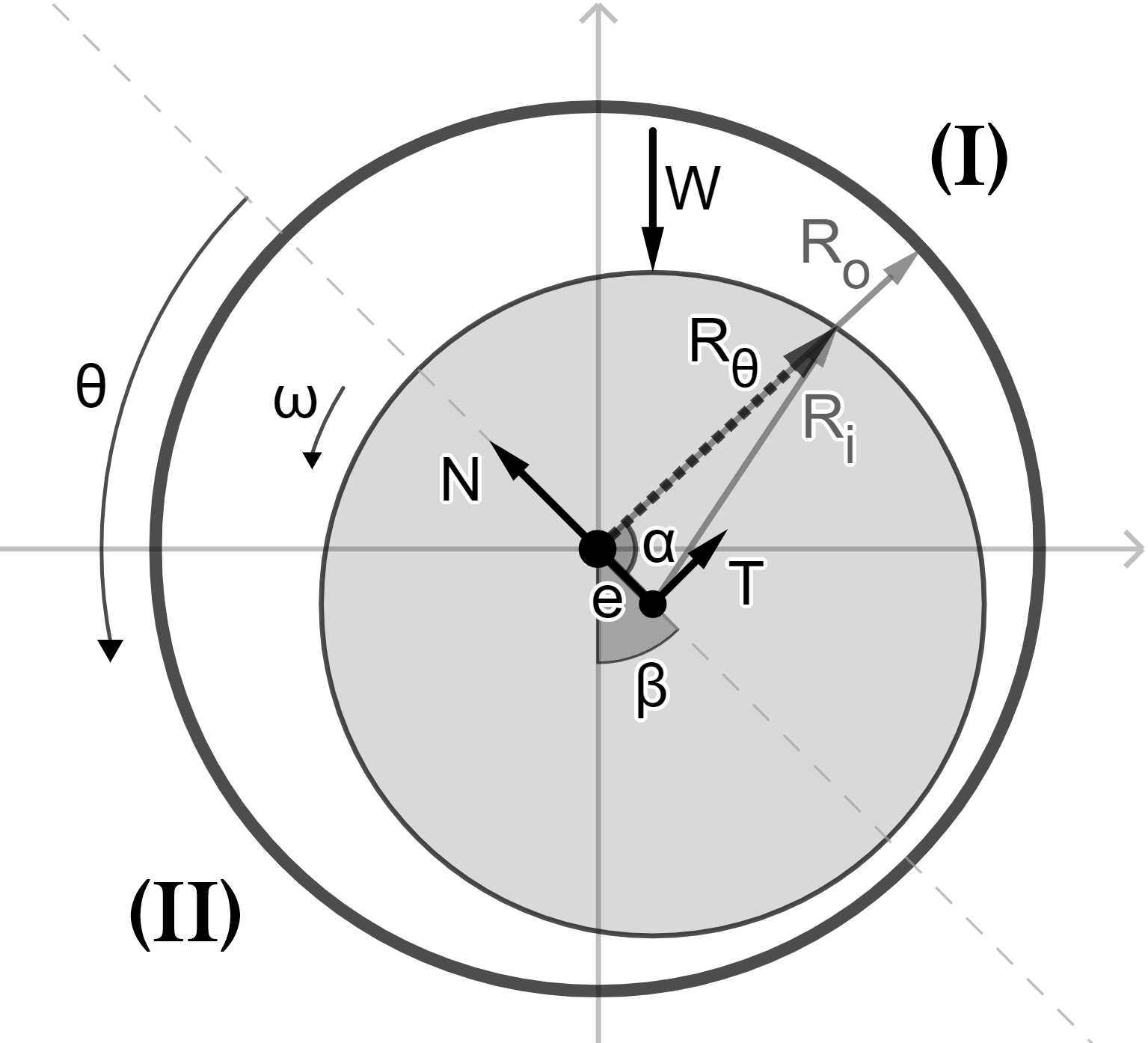}
    \caption{Schematic drawing of an eccentricity journal bearing} \label{fig:mancal}
\end{figure}

In this article, the dynamics of the fluid flow in the annular space between two cylinders is studied. This is a representation of a hydrodynamic bearing. The system is shown in Fig. \ref{fig:mancal}. The external cylinder, with radius $R_o$ is called stator and the internal cylinder, with radius $R_i$, is the rotor which spins with a speed $\omega$.

With the rotation speed, the rotor moves to an eccentric position. Any point on the rotor's surface can be described as (see Fig. \ref{fig:mancal})

\begin{align}\label{eq:rtheta}
    R_{\theta} = \sqrt{R_i ^2 - e^2 \sin^2{\alpha}} + e \cos{\alpha}\,,
\end{align}
with 
\begin{align}
 \alpha = 
\begin{cases} 
\dfrac{3\pi}{2} - \theta + \beta \text{,} 
&\mbox{if } \dfrac{\pi}{2} + \beta \leq \theta < \dfrac{3\pi}{2} + \beta \quad \textbf{(\textrm{I})} \\ \\
- \left(\dfrac{3\pi}{2} - \theta + \beta\right) \text{,} 
& \mbox{if } \dfrac{3\pi}{2} + \beta \leq \theta < \dfrac{5\pi}{2} + \beta \quad \textbf{(\textrm{II})}\nonumber
\end{cases},
\end{align}
where $e$ is the eccentricity, $\beta$ is the attitude angle, $R_o$ is the stator's radius and $\theta$ is the angle used to define $R_\theta$. Note that $\theta$ is measured from an axis passing through the points of maximum and minimum fluid width. Also, it is important to note that the difference between the radius of the stator and the rotor is very small compared to the radius of the stator.


\section{Theoretical Modeling}
The dynamic of the fluid between the rotor and stator is modeled using the Navier-Stokes equations and the continuity equation. For this problem, the equations are
\begin{subequations}
    \label{eq:NS_e_continuidade}
    \begin{align}
        \label{eq:NS_e_continuidade_A}
        \rho \Bigr(\frac{\partial \mathbf{v}}{\partial t} + \mathbf{v} \cdot \nabla  \mathbf{v}\Bigl) &= \nabla \cdot \sigma, \\
        \label{eq:NS_e_continuidade_B}
        \frac{\partial \rho}{\partial t} + \nabla \cdot (\rho \mathbf{v}) &= 0,
    \end{align}
\end{subequations}
where $ \rho $ is the specific mass of the fluid; $ \mathbf{v} $ is the velocity field vector in cylindrical coordinates with components $u$ (axial), $v$ (radial) and $w$ (tangential); and $\sigma = -p \mathbf{I} + \tau$ is Cauchy's tensor in terms of the pressure field $p$, the tension tensor $\tau$ and the identity tensor $\mathbf{I}$.


Assuming an incompressible Newtonian fluid in a steady-state laminar flow, Eqs. \eqref{eq:NS_e_continuidade} can be simplified \cite{Fox}.  First, $\rho$ can be considered constant. Also, the shear stress is proportional to its deformation rate, i.e., ${\mathbf{\tau}} = {\mathbf{\mu}} (\nabla \mathbf{v})$. Moreover, the fluid is considered in a steady state, where the quantities do not vary over time. Thus, Eq.  (\ref{eq:NS_e_continuidade}) can be rewritten as:
\begin{subequations}
    \label{eq:NS2}
        \begin{align}
        \label{eq:NS2_A}
        \rho (\mathbf{v} \cdot \nabla  \mathbf{v}) &= -\nabla p + \mu \nabla^2 \mathbf{v},\\
        \label{eq:NS2_B}
        \nabla & \cdot  \mathbf{v} = 0\,,
    \end{align}
\end{subequations}



which are expressed in cylindrical coordinates to preserve the curvature effects, following \cite{Pina:05,selma:99}. In most references it is assumed that the radius of curvature of the boundaries is large compared to the thickness of the lubricating film, and these effects are neglected. Developing Eq. \eqref{eq:NS2} yields the following equations.

\begin{subequations}
\label{eq:N-S}
\noindent \text{Axial direction, } z:
 \begin{align}\label{eq: N-S_cilindrica-z}
{\rho 
\left(
u \dfrac{\partial{u}}{\partial{z}} 
+ v \dfrac{\partial{u}}{\partial{r}} 
+ \dfrac{w}{r} \dfrac{\partial{u}}{\partial{\theta}}
\right)}
=
{-\dfrac{\partial{p}}{\partial{z}} 
+ \mu 
\left(
\dfrac{1}{r} \dfrac{\partial{}}{\partial{r}}\left[r\dfrac{\partial{u}}{\partial{r}} \right] 
+ \dfrac{1}{r^2}\dfrac{\partial^2{u}}{\partial{\theta ^2}} 
+ \dfrac{\partial^2{u}}{\partial{z^2}}  
\right)}\text{.} 
\end{align}
\noindent\text{Radial direction, } r:
\begin{align}\label{eq: N-S_cilindrica-r}
{
\rho 
\left(
v \dfrac{\partial{v}}{\partial{r}} 
+ u \dfrac{\partial{v}}{\partial{z}} 
+ \dfrac{w}{r} \dfrac{\partial{v}}{\partial{\theta}} 
- \dfrac{w^2}{r} \right)}
&=\nonumber\\
{{-\dfrac{\partial{p}}{\partial{r}} 
+ \mu 
\left(
\dfrac{\partial}{\partial r} \left[\dfrac{1}{r}\dfrac{\partial(r v)}{\partial r} \right] 
+ \dfrac{1}{r^2}\dfrac{\partial^2{v}}{\partial{\theta ^2}} 
- \dfrac{2}{r^2}\dfrac{\partial{w}}{\partial{\theta}}
+ \dfrac{\partial^2{v}}{\partial{z^2}} \right)}}\text{.} 
\end{align}
\text{Tangent direction, }$\theta$:
\begin{align}\label{eq: N-S_cilindrica-theta}
{\rho 
\left(
v \dfrac{\partial{w}}{\partial{r}} 
+ u \dfrac{\partial{w}}{\partial{z}} 
+ \dfrac{w}{r} \dfrac{\partial{w}}{\partial{\theta}} 
+ \dfrac{v w}{r} \right)}
&=\nonumber\\
{{-\dfrac{1}{r} \dfrac{\partial{p}}{\partial{\theta}}
+ \mu 
\left(
\dfrac{\partial}{\partial r} \left[\dfrac{1}{r}\dfrac{\partial(r w)}{\partial r} \right] 
+ \dfrac{1}{r^2}\dfrac{\partial^2{w}}{\partial{\theta ^2}} 
+ \dfrac{2}{r^2}\dfrac{\partial{v}}{\partial{\theta}}
+ \dfrac{\partial^2{w}}{\partial{z^2}} \right)}}\text{.}
\end{align}
\text{Continuity:}
\begin{align}\label{eq:continuity}
\dfrac{1}{r} \dfrac{\partial{\left(rv\right)}}{\partial{r}}+\dfrac{1}{r}\dfrac{\partial{w}}{\partial{\theta}}+\dfrac{\partial{u}}{\partial{z}} = 0\text{.}
\end{align}
\end{subequations}

This system of partial differential equations is nonlinear, and the velocity and pressure fields are unknown. After applying the proper boundary conditions, it can be approximated numerically. However,  additional simplifications can be applied when a dimensional analysis is performed.

\subsection{Dimensionless Navier-Stokes equations}


The fact that the radial clearance $F=R_o - R_i$ is very small compared to the radii $R_o$ and $R_i$, and the length $L$ is used to decide which terms can be neglected. First, consider $U$ a typical speed and $L$ a characteristic length with the same order of magnitude as $R_o$ and $R_i$. Then, the following dimensionless variables are introduced:
\begin{align}\label{eq:dimVars}
    u = U\hat{u}, \quad v = U\hat{v}, \quad w = U\hat{w},  \quad p = P\hat{p}, \quad
    P=\dfrac{\mu UL}{F^2},  \nonumber\\
    \quad \Delta z = L \Delta \hat{z},  \quad  \Delta r = F \Delta \hat{r},  \quad \Delta (r \theta) = L \Delta (\hat{r}{\theta})\,,
\end{align}

where the hat symbol means dimensionless variable. Note that the expression for $P$ was used since the pressure exhibits a higher variation near the smallest fluid width. Now, using the dimensionless variables \eqref{eq:dimVars} in (\ref{eq: N-S_cilindrica-z}) yields:



\begin{align}
\begin{split}\label{eqzADM}
     & \rho \left( U\hat{u}\pdv{U\hat{u}}{L\hat{z}} + U\hat{v}\pdv{U\hat{u}}{F\hat{r}} + \frac{U\hat{w}}{L\hat{r}}\pdv{U\hat{u}}{\theta} \right) = \\
    &  - \pdv{P\hat{p}}{L\hat{z}} + \mu \left(\frac{1}{L\hat{r}}\pdv{}{F\hat{r}}\left[ L\hat{r}\pdv{U\hat{u}}{
    F\hat{r}}\right] + \frac{1}{L^2 \hat{r}^2}\pdv[2]{U\hat{u}}{\theta} + \pdv[2]{U\hat{u}}{L^2 \hat{z}} \right)\,.
\end{split}
\end{align}

After some algebraic manipulations one obtains
\begin{align}
\begin{split}\label{eqzRe}
     & Re   \left( \left( \frac{F^2}{L^2} \right) \hat{u}\pdv{\hat{u}}{\hat{z}} + \left(\frac{F}{L}\right)\hat{v}\pdv{\hat{u}}{\hat{r}} + \left( \frac{F^2}{L^2} \right) \frac{\hat{w}}{\hat{r}}\pdv{\hat{u}}{\theta} \right) = \\
    &   - \pdv{\hat{p}}{\hat{z}} +  \left(\frac{1}{\hat{r}}\pdv{}{\hat{r}}\left[ \hat{r}\pdv{\hat{u}}{
    \hat{r}}\right] + \left( \frac{F^2}{L^2} \right) \frac{1}{ \hat{r}^2}\pdv[2]{\hat{u}}{\theta} + \left( \frac{F^2}{L^2} \right) \pdv[2]{\hat{u}}{ \hat{z}} \right),
\end{split}
\end{align}

where the Reynolds number
\begin{align}
    Re = \frac{\rho U L}{\mu}
\end{align}
is used.
Following the traditional theory of lubrication, the terms multiplied by $(F / L)^2 $ and $ (F / L) $ are neglected and Eqs. \eqref{eq: N-S_cilindrica-z}, \eqref{eq: N-S_cilindrica-r} and \eqref{eq: N-S_cilindrica-theta} become:
\begin{subequations}
    \begin{align} \label{eq:simpz}
          - \pdv{p}{z} +  \frac{1}{r}\pdv{}{r}\left(r\pdv{u}{r}\right) &= 0,\\
    \label{eq:simpr}
         -\pdv{p}{r} &= 0,\\
    \label{eq:simptheta}
        \pdv{}{r}\left(\frac{1}{r}\pdv{(r w)}{r}\right) - \frac{1}{r}\pdv{p}{\theta} &= 0.
    \end{align}
\end{subequations}

After this process, the number of terms was reduced considerably. For the equation along the radial direction \eqref{eq:simpr}, the radial velocity disappeared and the pressure gradient along this direction is zero. 

\subsection{Velocities}

Equations (\ref{eq:simpz}) and (\ref{eq:simptheta}) can now be integrated in order to compute the velocities $u$ and $w$:

\begin{subequations}
    \begin{equation}
        u = \left( \pdv{p}{z}  \right)\frac{1}{4}r^2 + c_1 \ln r + c_2\,,
    \end{equation}
    \begin{equation}
        w = \frac{1}{2 }\pdv{p}{\theta}r\left(\ln r-\frac{1}{2}\right) + c_3 r + \frac{c_4}{r}\,,
    \end{equation}
\end{subequations}

where $ c_1 $, $ c_2 $, $ c_3 $ and $ c_4 $ are constants. Since the fluid speed is zero at the stator and equal to the tangential speed $W$ at the rotor's surface, the boundary conditions are: $u(R_o) = 0$, $u(R_{\theta})=0$, $w(R_o)=0$ and $w(R_{\theta}) = \omega R_{i} = W$. Thus, the description of the velocities $ u $ and $ w $ is obtained as a function of the pressure gradient:
\begin{subequations}
    \begin{align} 
        \label{u}
         u = &\pdv{p}{z} \frac{R^2_{\theta}}{4}\left[\left(\frac{r}{R_{\theta}}\right)^2-\frac{(R^2_o-R^2_{\theta})}{R^2_i\ln(R_o/R_{\theta})}\ln\left(\frac{r}{R_{\theta}}\right)-1\right],\\
        w &=\dfrac{1}{2} \dfrac{\partial p}{\partial \theta} \left[ r \left( \ln r - \dfrac{1}{2}\right) + K r  - \dfrac{R_o^2}{r} \left( \ln R_o + K - \dfrac{1}{2}\right)\right] \nonumber\\
        &+ \dfrac{W R_{\theta}}{\left(R_{\theta}^2 - R_o^2\right)} \left(r - \dfrac{R_o^2}{r}\right), 
        \label{w}
    \end{align}
\end{subequations}
where
\begin{equation}
    K = \frac{1}{R^2_o-R^2_{\theta}}\left[R^2_{\theta}\left(\ln R_{\theta} - \frac{1}{2}\right)-R^2_o\left(\ln R_o - \frac{1}{2}\right)\right].
\end{equation}

\subsection{Continuity Equation}

In order to obtain the pressure field, the continuity equation (\ref{eq:continuity}) is integrated in the annular region of interest,
 
\begin{equation} \label{eq:cont-integrate}
    \underset{\mathrm{I}}{\underbrace{
        \int^{R_{o}}_{R_{\theta}} 
        \left(
            \frac{\partial{(rv)}}{\partial{r}}
        \right) \,dr
        }}
    + \underset{\mathrm{II}}{\underbrace{
        \int^{R_{o}}_{R_{\theta}}
        \left(
            \frac{\partial{w}}{\partial{\theta}}
        \right) \,dr
        }}
    + \underset{\mathrm{III}}{\underbrace{
        \int^{R_{o}}_{R_{\theta}}
        \left(
            \frac{\partial{(ru)}}{\partial{z}}
        \right) \,dr
    }}
    = 0\,.
\end{equation}
Each term is detailed as follows:
\begin{align} \nonumber
    &(\mathrm{I}) \quad \int^{R_{o}}_{R_{\theta}} 
        \left(
            \frac{\partial{(rv)}}{\partial{r}}
        \right) \,dr 
        = 
        R_{o} v(R_{o}) - R_{\theta} v(R_{\theta})\,,
    \\ \nonumber
    &(\mathrm{II}) \quad \int^{R_{o}}_{R_{\theta}}
        \left(
            \frac{\partial{w}}{\partial{\theta}}
        \right) \, dr 
        =
        \dfrac{\partial}{\partial \theta}
        \int^{R_{o}}_{R_{\theta}} w \,dr 
        - \left[
            w(R_{o}) \dfrac{\partial R_{o}}{\partial \theta}
            - w(R_{\theta}) \dfrac{\partial R_{\theta}}{\partial \theta}
        \right]\,,
    \\ \nonumber
    &(\mathrm{III}) \quad \int^{R_{o}}_{R_{\theta}}
        \left(
            \frac{\partial{(ru)}}{\partial{z}}
        \right) \, dr
        =
        \dfrac{\partial}{\partial z}
        \int^{R_{o}}_{R_{\theta}} (ru) \,dr 
        - \left[
            u(R_{o}) \dfrac{\partial R_{o}}{\partial z}
            - u(R_{\theta}) \dfrac{\partial R_i}{\partial z}
        \right]\,.
\end{align}

\begin{figure}  
    \centering
    \includegraphics[width=.5\textwidth]{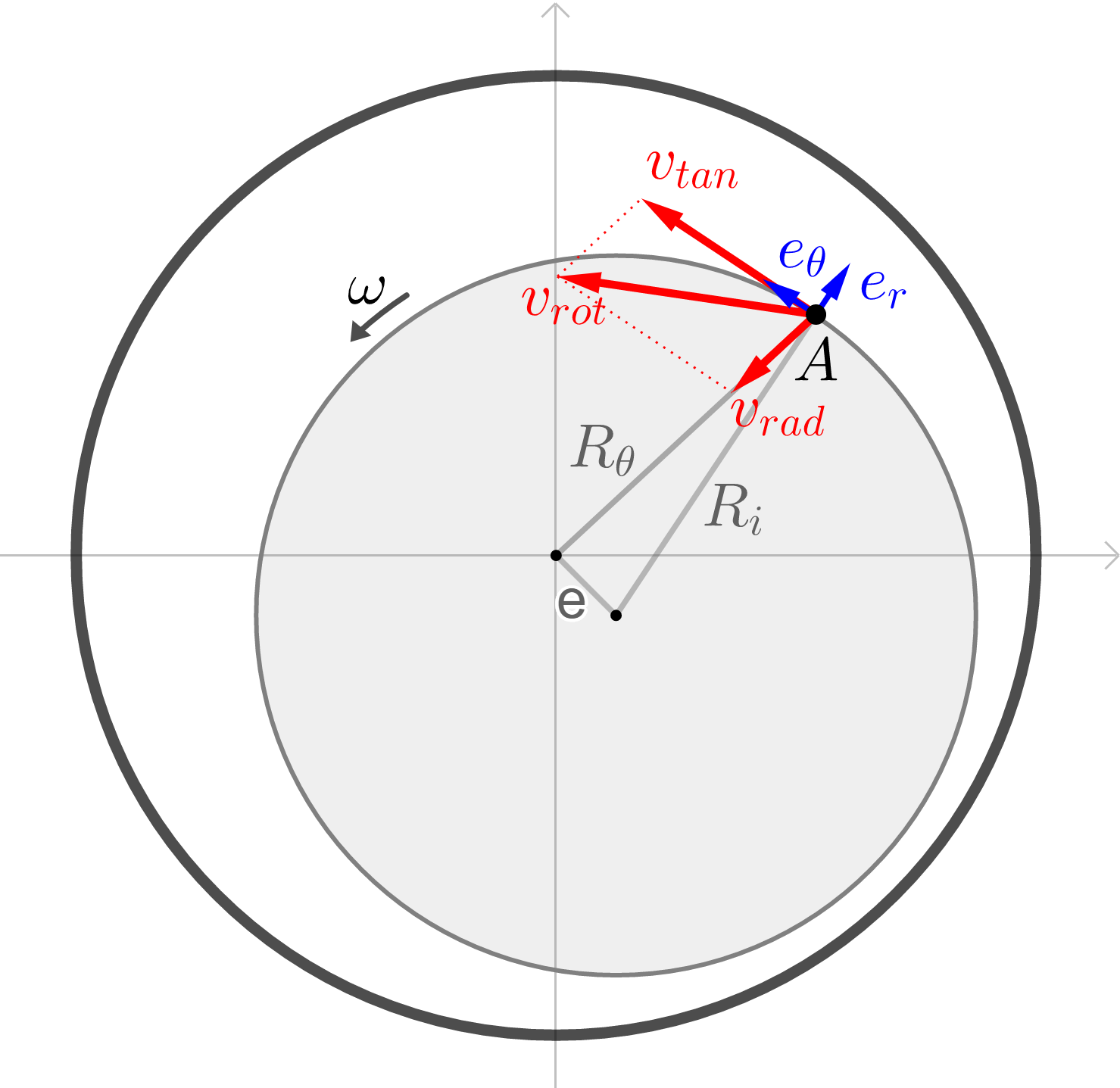}
    \caption{Graphical description of the speed of point A.} \label{fig:integral}
\end{figure}

Some remarks about the boundary conditions should be registered. The radial speed at the stator  $v(R_{o})$ is zero; although the speed at the rotor's surface is tangential to it, it has a projection along the radial direction in the coordinate system with origin at the stator's center (see Fig. \ref{fig:integral}); since the external radius does not vary along the tangential direction, $ \dfrac {\partial R_o} {\partial \theta}$ is zero; due to the eccentricity, $ \dfrac {\partial R_{\theta}} {\partial \theta}$ is not zero; the speeds $u(R_o)$ and $u(R_\theta)$ are zero.


After applying the aforementioned boundary conditions, Eq. (\ref{eq:cont-integrate}) is rewritten as:

\begin{align} \label{eq:int-continuidade-simplif_1}
    \dfrac{\partial}{\partial \theta}
    \int^{R_{o}}_{R_{\theta}} w \,dr
    + w(R_{\theta})\dfrac{\partial R_{\theta}}{\partial \theta}
    + \dfrac{\partial}{\partial z}
    \int^{R_{o}}_{R_{\theta}} ru \,dr
    - R_{\theta} v(R_{\theta})
    = 0
\end{align}
The remaining boundary conditions $w(R_\theta)$ and $v(R_\theta)$ are obtained by kinematics. First, consider a point $A$ at the surface of the rotor, as shown in Fig. \ref{fig:integral}. The poition vector of this point, with respect to a cylindrical coordinate system with origin at the stator's center, can be described as
\begin{align}\label{eq:pos-A}
    \vec{p}_A=R_\theta\vec{e}_\theta,
\end{align}
and its time derivative as
\begin{align}\label{eq:vel-A}
    \vec{v}_A= \dot{R_{\theta}}\,\vec{e}_r + R_{\theta}\,\dot{\vec{e}}_r
    = \underbrace{\omega \dfrac{\partial R_{\theta}}{\partial \theta}}_{v_{\text{rad}}}\,\vec{e}_r
    + \underbrace{\omega R_{\theta}}_{v_{\text{tan}}}\,\vec{e}_{\theta},
\end{align}
where
\begin{subequations}\label{eq:vtan-vrad}
   \begin{align}
    v_{\text{rad}}&=u(R_\theta)=\omega \dfrac{\partial R_{\theta}}{\partial \theta},\\
    v_{\text{tan}}&=w(R_\theta)=\omega R_{\theta}.
\end{align}
\end{subequations}


The boundary conditions in Eq. \eqref{eq:vtan-vrad} are substituted in Eq. (\ref{eq:int-continuidade-simplif_1}), which yields:

\begin{align} \label{eq:int-continuidade-simplif}
    \dfrac{\partial}{\partial \theta}
    \int^{R_{o}}_{R_{\theta}} w \,dr
    + \dfrac{\partial}{\partial z}
    \int^{R_{o}}_{R_{\theta}} ru \,dr
    = 0
\end{align}

Finally, the speeds $u$ and $w$ in Eqs. \eqref{u} and \eqref{w} are substituted in Eq.  \eqref{eq:int-continuidade-simplif} and integrated:

\begin{equation}\label{eq:reynolds}
    \pdv{}{\theta}\left(C_1\pdv{p}{\theta}\right)+\pdv{}{z}\left(C_2\pdv{p}{z}\right) = \pdv{}{\theta}C_{0}
\end{equation}
where:
\begin{equation}
    C_{0}=-WR_{\theta}\left[\ln\left(\frac{R_o}{R_{\theta}}\right)\left(1+\frac{R^2_{\theta}}{(R^2_o-R^2_{\theta})}\right)-\frac{1}{2}\right]
\end{equation}
\begin{align}
    \begin{split}
        C_1 = &\frac{R_{\theta}}{2}\left[\frac{1}{2R_{\theta}}[R^2_o\ln R_o - R^2_{\theta}\ln R_{\theta} - \left(R^2_o - R^2_{\theta}\right)\left(1+K\right)]\right]-\\
        &\frac{R^2_{\theta}}{2\mu}\left[\left(\ln R_{\theta} - \frac{1}{2}+K\right)\ln\left(\frac{R_o}{R_{\theta}}\right)\right]
    \end{split}
\end{align}
\begin{align}
    \begin{split}
        C_2 = &\frac{-R^2_{\theta}}{8}\left[R^2_o-R^2_{\theta}-\frac{(R^4_o-R^4_{\theta})}{2R^2_{\theta}}\right]+\\
        &\left( \frac{R^2_o-R^2_{\theta}}{R^2_r\ln(R_o/R_{\theta})}\right) \left[ R^2_o\ln\left(\frac{R_o}{R_{\theta}}\right)-\frac{(R^2_o-R^2_{\theta})}{2}\right] 
    \end{split}
\end{align}

The original Navier-Stokes equation, Eq. \eqref{eq:N-S}, was reduced to an elliptical differential equation, Eq. \eqref{eq:reynolds}, which is known as the Reynolds equation that can be approximated numerically.

\section{Numerical Approximation for the pressure field}
In this section, an approximate solution to Eq. \eqref{eq:reynolds} is obtained for the pressure field $p$. A central finite difference approximation method is used (see Ref. \cite{book:burden} for more details).

\subsection{Discretization of the domain}

\begin{figure} 
    \centering
    \includegraphics[width=.5\textwidth]{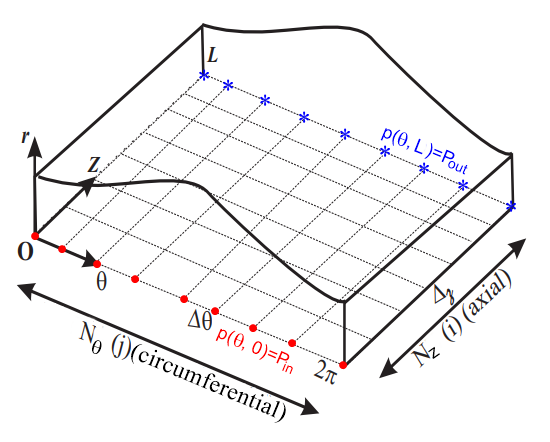}
    \includegraphics[width=.4\textwidth]{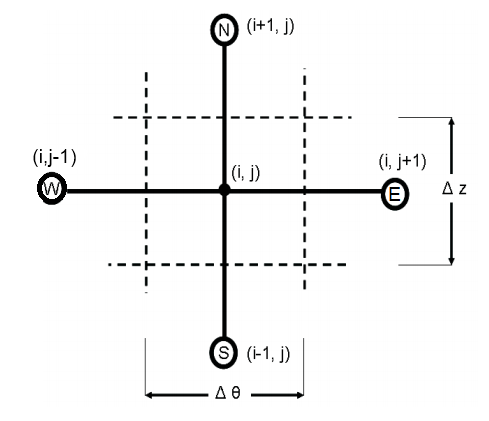}
    \caption{Rectangular mesh used for discretization of the equation and graphic translation of the centered finite difference method; adapted from \cite{selma:99}.}\label{fig:Mash}
\end{figure}

The cylindrical region of interest is opened up and the discretization is performed in a rectangular region with $N_z \times N_\theta$ nodes. Thus, the fluid film has $N_z$ divisions along the axial direction and $N_z$ division along the tangential direction, both equally spaced. Fig. \ref{fig:Mash} shows this procedure.


\subsection{Central finite differences formula}
The partial derivatives of Eq. \eqref{eq:reynolds} are dicretized according to the following formulas:
\begin{align}\label{eq:finite-diff}
    \pdv{f(i,j)}{\theta} = \frac{f(i,j-1) - f(i,j+1)}{2\Delta\theta}  \quad , \quad
    \pdv{f(i,j)}{z} = \frac{f(i-1,j) - f(i+1,j)}{2\Delta z} 
\end{align}

After the substitution of Eq. \eqref{eq:finite-diff} in Eq. \eqref{eq:reynolds}, the following algebraic equation is obtained:

\begin{align} \label{eq:P}
    &p_{i-1,j}\frac{(C_{2(i-1,j)})}{\Delta z^{2}} + \nonumber  p_{i,j-1}\frac{(C_{1(i,j-1)})}{\Delta\theta^{2}} - p_{i,j}\left(\frac{(C_{1(i,j)} + C_{1(i,j-1)})}{\Delta\theta^{2}} + \frac{(C_{1(i,j)} + C_{1(i,j-1)})}{\Delta z^{2}} \right) \nonumber  + \\
    &p_{i,j+1} \frac{(C_{1(i,j)})}{\Delta\theta^{2}} + 
    p_{i+1,j}\frac{(C_{2(i,j)})}{\Delta z^{2}} = \frac{1}{\Delta\theta}\left[C_{0W(i,j)}-C_{0W(i,j-1)}\right],
\end{align}

where the pressure at the node $(i,j)$ is written as $p(i,j)$. The boundary conditions of Eq. \eqref{eq:P} are:
\begin{align*}
    & p(z=0)=p(1,j)= P_{in},\\
    & p(z=L)=p(N_Z,j)=P_{out},\\
    & p(\theta=0)=p(\theta = 2\pi) =  p(i,1)=p(i,N_\theta).
\end{align*}
Eq. \eqref{eq:P} can be expressed in a matrix form as 
\begin{align}\label{eq:Mp-f}
    \mat{M}\vec{p}=\vec{f}\,,
\end{align}
where
\begin{align}
    \vec{p}
    =
    \begin{bmatrix}
        p_{0,0}&p_{0,1}&\dots&p_{N_Z,N_\theta}
    \end{bmatrix}^T,\\
    \vec{f}
    =
    \begin{bmatrix}
        f_{0,0}&f_{0,1}&\dots&f_{N_Z,N_\theta}
    \end{bmatrix}^T,
\end{align}
and $\mat{M}$ is a sparse matrix. Eq. \eqref{eq:Mp-f} is assembled in \textit{Python} and solved with the SymPy\footnote{\url{https://docs.scipy.org/doc/scipy/reference/generated/scipy.sparse.linalg.spsolve.html}} library.

\section{Fluid Film Forces}

Integrating the pressure field one obtains the fluid film force that is applied to the rotor. The decomposition of this force in the orthogonal axis N-T shown in Fig. \ref{fig:mancal} are given by \cite{Yamamoto}:


\begin{subequations}
 \begin{align}
    f_N &= -R_i\int_{0}^{L}\int_{0}^{\pi}p(\theta,z)\cos{\theta}d\theta dz \\
    f_T &= R_i\int_{0}^{L}\int_{0}^{\pi}p(\theta,z)\sin{\theta}d\theta dz
\end{align}  \label{eq:force}
\end{subequations}

or in a X-Y axis by:
\begin{subequations}
\begin{align}
    f_x = f_T \cos{(\beta)} - f_N \sin{(\beta)}\\
    f_y = f_T \sin{(\beta)} + f_N \cos{(\beta)}
\end{align}
\end{subequations}

Since the pressure field is discrete, the double integrals are computed numerically using the Simpson's method.

\subsection{Equilibrium position}
During the normal operation of a rotating machine supported by hydrodynamic bearings, external loads are applied to the rotor and the fluid film forces oppose to them. If only the weight of the rotor is considered, the center of the spinning rotor stays in an equilibrium position where the weight is balanced with the fluid film forces. This position can be described by the eccentric displacement $e$ and the attitude angle $\beta$, as shown in Fig. \ref{fig:mancal}. At this position, the following condition must be satisfied:
\begin{subequations}\label{eq:equilibrio}
    \begin{align}
        f_{x_0} =& 0, \\
        f_{y_0} =& W, 
    \end{align}
\end{subequations}
where $W$ is the weight of the rotor.

An iterative method is used to find the position of the rotor that meets the condition of Eq. \eqref{eq:equilibrio}. Starting with an initial (guess) position, the difference between the current forces and the forces shown in Eq. \eqref{eq:equilibrio} is calculated. Next, the rotor center is slightly moved, and the forces are recalculated. When a tolerance condition is met, the method stops. More details can be found in the \textit{Python library} \textit{optimize.least\_squares} 

The static performance characteristics of the bearing are listed in a nondimensional number known as Sommerfeld Number ($S$) \cite{SanAndres4}:
\begin{align}\label{eq:sommerfeld}
    S = \frac{\mu \omega L R_o^3}{\pi W F^2}\text{,}
\end{align}
where $F=R_o - R_i$ is the radial clearance.

\begin{figure}[ht]
    \centering
    \includegraphics[width=.4\textwidth]{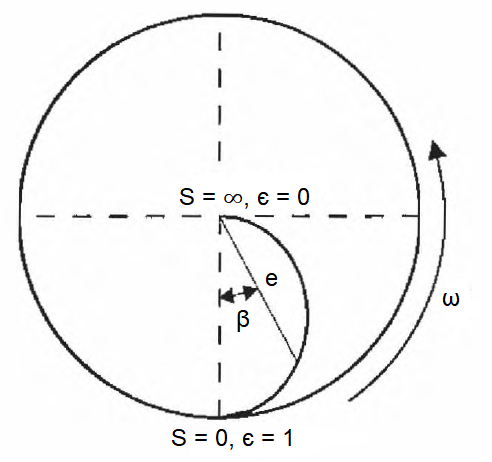}
    \caption{Path taken by the center of the rotor related to the Sommerfeld Number; adapted from \cite{Friswell}.}
    \label{fig:sommerfeld}
\end{figure}

This nondimensional number is directly linked to the equilibrium position. Fig. \ref {fig:sommerfeld} illustrates this relation.

\section{Dynamic Coefficients}

In the previous section, the fluid film forces calculated are valid only at the static equilibrium of the rotor's center. If the rotor performs small movements around such position, the fluid film forces vary. One way to characterize the relation between the forces and small displacements around the equilibrium position is to determine the stiffness and damping coefficients.

\subsection{Radial and tangential speeds}
Here, it is considered that the rotor is not only spinning but also moving along the horizontal / vertical directions. Thus, the speeds $w(R_\theta)$ and $v(R_\theta)$ in the continuity Eq.  \eqref{eq:int-continuidade-simplif_1} must be determined with this new dynamic state.

Let us consider again a point $A$ belonging to the surface of the rotor. If the rotor is spinning and the velocity of its center is being perturbed, its total velocity is
\begin{align}\label{eq:vtot}
    \vec{v}_A = \vec{v}_r + \vec{v}_p
\end{align}
where $\vec{v}_r$ is the velocity of $A$ due to the rotation speed and $\vec{v}_p$ is the velocity acquired by the perturbation. In order to derive an equation for $\vec{v}_p$, the small displacements around the equilibrium position $(x_0,y_0)$ are assumed of the following form:
\begin{subequations}\label{eq:pert-x}
   \begin{align}
    x = x_0 + x_p \sin (\omega_p t),
    \\
    y = y_0 + y_p \sin (\omega_p t),
\end{align} 
\end{subequations}
where $x_p$ and $y_p$ are the maximum amplitudes of the perturbations and $\omega_p$ is the perturbation frequency along each axis. The time derivative of Eq. \eqref{eq:pert-x} gives the two terms of $\vec{v}_p$:
\begin{subequations}\label{eq:pert-v}
   \begin{align}
    (\vec{v}_p)_x = \omega_p x_p \cos (\omega_p t)\vec{e}_x,
    \\
    (\vec{v}_p)_y = \omega_p y_p \cos (\omega_p t)\vec{e}_y,
\end{align} 
\end{subequations}
where $(\vec{v}_p)_x$ and $(\vec{v}_p)_y$ stand for the speeds when perturbations along the horizontal and vertical directions are applied, respectively. In cylindrical coordinates, and using Eq. \eqref{eq:pert-v}, one arrives to:





\begin{align}\label{eq:vtot-x-y}
    (\vec{v}_{A})_x 
    = \underset{(v_{\text{rad}})_x}{\underbrace{\left(\omega \dfrac{\partial R_{\theta}}{\partial \theta} + \omega_p x_p \cos{(\omega_p t)\cos{(\theta)}}\right)}} \vec{e}_r 
    + \underset{(v_\text{tan})_x}{\underbrace{\left(\omega R_{\theta} - \omega_p x_p \cos{(\omega_p t)}\sin{(\theta)} \right)}} \vec{e}_{\theta}\nonumber
    \\
    (\vec{v}_{A})_y 
    = \underset{(v_{\text{rad}})_y}{\underbrace{\left(\omega \dfrac{\partial R_{\theta}}{\partial \theta} + \omega_p y_p \cos{(\omega_p t)\sin{(\theta)}}\right)}} \vec{e}_r 
    + \underset{(v_{\text{tan}})_y}{\underbrace{\left(\omega R_{\theta} - \omega_p y_p \cos{(\omega_p t)}\cos{(\theta)} \right)}} \vec{e}_{\theta}
\end{align}
where
\begin{subequations}\label{eq:vtan-vrad-2}
   \begin{align}
    (v_\text{rad})_x&=u(R_\theta)_x
    =
    \omega \dfrac{\partial R_{\theta}}{\partial \theta} + \omega_p x_p \cos{(\omega_p t)\cos{(\theta)}},\\
    (v_\text{tan})_x&=w(R_\theta)_x
    =
    \omega R_{\theta} - \omega_p x_p \cos{(\omega_p t)}\sin{(\theta)} ,\\
    (v_\text{rad})_y&=u(R_\theta)_y
    =
    \omega \dfrac{\partial R_{\theta}}{\partial \theta} + \omega_p y_p \cos{(\omega_p t)\sin{(\theta)}},\\
    (v_\text{tan})_y&=w(R_\theta)_y
    =
    \omega R_{\theta} - \omega_p y_p \cos{(\omega_p t)}\cos{(\theta)}.
    \end{align}
\end{subequations}

\subsection{Additional terms in the continuity equation}

Equations \eqref{eq:vtan-vrad-2}a-d are used as the new boundary conditions in the continuity equation, Eq. \eqref{eq:int-continuidade-simplif_1}. This consideration affects only the coefficient $C_0$ of Eq. \eqref{eq:int-continuidade-simplif_1}, which is
\begin{subequations}
   \begin{align}\label{eq:new_c0}
    (C_0)_x &= C_0 + R_{\theta} \omega_p x_p \cos{(\omega_p t)} \sin{(\theta)},
    \\
    (C_0)_y &= C_0 + R_{\theta} \omega_p y_p \cos{(\omega_p t)} \cos{(\theta)}.
    \end{align}
\end{subequations}

Thus, for an excitation along the $x$ direction, the fluid film forces are computed according to the equation:
\begin{subequations}\label{eq:perturbed-forces}
\begin{align}
    (f_x(t))_x = &R_i\!\!\!\int_{0}^{L}\!\!\!\!\int_{0}^{\pi}\!\!\!\! p(\theta,z,t;x_p,\omega_p)\sin{\theta}\cos{\beta}d\theta dz  \nonumber \\ 
    & \hspace{3cm} +R_i\!\!\!\int_{0}^{L}\!\!\!\!\int_{0}^{\pi}\!\!\!\!p(\theta,z,t;x_p,\omega_p)\cos{\theta} \sin{\beta} d\theta dz \\
    (f_y(t))_x = &R_i\!\!\!\int_{0}^{L}\!\!\!\!\int_{0}^{\pi}\!\!\!\!p(\theta,z,t;x_p,\omega_p)\sin{\theta}\sin{\beta}d\theta dz  \nonumber\\
    & \hspace{3cm} -R_i\!\!\!\int_{0}^{L}\!\!\!\!\int_{0}^{\pi}\!\!\!\! p(\theta,z,t;x_p,\omega_p)\cos{\theta} \cos{\beta}d\theta dz,
\end{align}
\end{subequations}
where $p(\theta,z,t;x_p,\omega_p)$ is the pressure field computed at time $t$ with perturbation amplitude and frequency $x_p$ and $\omega_p$, respectively. Similar equations are valid for a perturbation along the $y$ direction.
\subsection{Stiffness and damping coefficients}

The forces of the fluid film can be considered general functions of the displacements and speeds of the center of the rotor. Thus, if small displacements around the center are applied, the expansions of the fluid forces in Taylor series yield
\begin{subequations}\label{eq:taylor}
   \begin{align}
    f_x &= f_{x_0} 
    + K_{xx}\Delta x
    + K_{xy} \Delta y
    + C_{xx} \Delta \dot{x}
    + C_{xy} \Delta \dot{y},
     \\
    f_y &= f_{y_0} 
    + K_{yx} \Delta x
    + K_{yy} \Delta y
    + C_{yx} \Delta \dot{x}
    + C_{yy} \Delta \dot{y}.
    \end{align}
\end{subequations}
where 
\begin{align}
    K_{xx}=\pdv{f_x}{x},K_{xy}=\pdv{f_x}{y},\nonumber\\
    K_{yx}=\pdv{f_y}{x},K_{yy}=\pdv{f_y}{y},\nonumber\\
    C_{xx}=\pdv{f_x}{\dot{x}},C_{xy}=\pdv{f_x}{\dot{y}},\nonumber\\
    C_{yx}=\pdv{f_y}{\dot{x}},C_{yy}=\pdv{f_y}{\dot{y}}.\nonumber
\end{align}

The terms $K_{xx}$, $K_{yy}$, $C_{xx}$ and $C_{yy}$ are called the direct stifness coefficients and direct damping coefficients, respectively. $K_{xy}$, $K_{yx}$, $C_{xy}$ and $C_{yx}$ are called the cross-coupled stiffness coefficients and cross-coupled damping coefficients, respectively.

By including the effects of the perturbation in the simplified equation \eqref{eq:reynolds}, it is possible to rewrite the Taylor Series expansion of the forces in Eq. \eqref{eq:taylor} as a function of the  time $t$. Thus, for each direction of disturbance the fluid film forces are:
\begin{align}\label{eq:taylor-com-tempo}
    \left( f_x(t) \right)_x &= f_{x_0} + k_{xx}\Delta x(t) + c_{xx}\Delta \dot{x}(t)\text{,}\nonumber
    \\
    \left( f_y(t) \right)_x &= f_{y_0} + k_{yx}\Delta x(t) + c_{yx}\Delta \dot{x}(t)\text{,}\nonumber
    \\
    \left( f_x(t) \right)_y &= f_{x_0} + k_{xy}\Delta y(t) + c_{xy}\Delta \dot{y}(t)\text{,}\nonumber
    \\
    \left( f_y(t) \right)_y &= f_{y_0} + k_{yy}\Delta y(t) + c_{yy}\Delta \dot{y}(t).
\end{align}
Note that the $\Delta x(t)$, $\Delta \dot{x}(t)$, $\Delta y(t)$ and $\Delta \dot{y}(t)$ are known by Eq. \eqref{eq:pert-x}; $f_{x0}$ and $f_{y0}$ are the forces at the equilibrium position from Eq. \eqref{eq:equilibrio}; and $(f_x(t))_x$, $(f_y(t))_x$, $(f_x(t))_y$ and $(f_y(t))_y$ are obtained from Eq. \eqref{eq:perturbed-forces}.

Equation (\ref{eq:taylor-com-tempo}) can be written in matrix form as:

\begin{itemize}
    \item Towards $x$:
\end{itemize}
\begin{align}\label{eq:matrix-coeficientes-x}
    \underset{\mathbf{A}_x}{\underbrace{
    \begin{bmatrix}
    	\Delta x[0] & \Delta \dot{x}[0]\\
    	\Delta x[1] & \Delta \dot{x}[1]\\
    	\vdots & \vdots \\
    	\Delta x[N-1] & \Delta \dot{x}[N-1]
	\end{bmatrix}
	}}
	\begin{bmatrix}
    	k_{xx} & k_{yx}\\
    	c_{xx} & c_{yx}
	\end{bmatrix}
	=
	\underset{\mathbf{F}_x}{\underbrace{
	\begin{bmatrix}
	(f_x[0])_x  - f_{x_0} & (f_y[0])_x - f_{y_0}\\
	(f_x[1])_x  - f_{x_0} & (f_y[1])_x - f_{y_0}\\
	\vdots & \vdots\\
	(f_x[N-1])_x  - f_{x_0} & (f_y[N-1])_x - f_{y_0}
	\end{bmatrix}}}\text{.}
\end{align}

\begin{itemize}
    \item Towards $y$:
\end{itemize}
\begin{align}\label{eq:matrix-coeficientes-y}
	\underset{\mathbf{A}_y}{\underbrace{
    \begin{bmatrix}
    	\Delta y[0] & \Delta \dot{y}[0]\\
    	\Delta y[1] & \Delta \dot{y}[1]\\
    	\vdots & \vdots \\
    	\Delta y[N-1] & \Delta \dot{y}[N-1]
	\end{bmatrix}
	}}
	\begin{bmatrix}
    	k_{xy} & k_{yy}\\
    	c_{xy} & c_{yy}
	\end{bmatrix}
	=
	\underset{\mathbf{F}_y}{\underbrace{
	\begin{bmatrix}
	(f_x[0])_y  - f_{x_0} & (f_y[0])_y - f_{y_0}\\
	(f_x[1])_y  - f_{x_0} & (f_y[1])_y - f_{y_0}\\
	\vdots & \vdots\\
	(f_x[N-1])_y  - f_{x_0} & (f_y[N-1])_y - f_{y_0}
	\end{bmatrix}}}\text{.}
\end{align}

Or, in order to simplify the notation:
\begin{align}\label{eq:matrix-coeficientes2}
    \mathbf{A}_x 
    \begin{bmatrix}
    	k_{xx} & k_{yx}\\
    	c_{xx} & c_{yx}
	\end{bmatrix}
	= \mathbf{F}_x 
	\quad \quad \text{e} \quad \quad
	\mathbf{A}_y 
    \begin{bmatrix}
    	k_{xy} & k_{yy}\\
    	c_{xy} & c_{yy}
	\end{bmatrix}
	= \mathbf{F}_y\text{.}
\end{align}

In order to calculate the values of the coefficients, the Least Squares method is employed. For that, one uses the Moore-Penrose pseudo inverse of the matrices $\mathbf{A}_x$ and $\mathbf{A}_y$.
\begin{align}\label{eq:moore-penrose}
    \mathbf{A}_{x}^{+} = 
    \left( \mathbf{A}_{x}^{T} \mathbf{A}_{x}\right)^{-1}
    \mathbf{A}_{x}^{T}
    \quad \quad \text{e} \quad \quad
    \mathbf{A}_{y}^{+} = 
    \left( \mathbf{A}_{y}^{T} \mathbf{A}_{y}\right)^{-1}
    \mathbf{A}_{y}^{T}\text{.}
\end{align}

And finally, the stiffness and damping coefficients are obtained:
\begin{align}\label{eq:coeficientes}
    \begin{bmatrix}
    	k_{xx} & k_{yx}\\
    	c_{xx} & c_{yx}
	\end{bmatrix}
	= \mathbf{A}_{x}^{+} \mathbf{F}_x \quad \quad \text{e} \quad \quad
    \begin{bmatrix}
    	k_{xy} & k_{yy}\\
    	c_{xy} & c_{yy}
	\end{bmatrix}
	= \mathbf{A}_{y}^{+}\mathbf{F}_y\text{.}
\end{align}



\section{Results and Discussion}


The choice of parameterization of the model presented in this work allows easy adaptation to different geometries and operational conditions of rotating machine components. This section analyzes three different cases: (1) short journal bearing, (2) elliptical bearing, and (3) worn bearing.


The algorithm used was developed in \textit{\textit{Python}}, and is available on the \textit{GitHub} platform, through the software \textit{Ross-Rotordynamics}\footnote{\url{https://github.com/ross-rotordynamics/}} \cite{Ross}, an open source library for rotodynamic analysis.

\subsection{Short journal bearing}

In the literature, it is common to use the Reynolds equation approximated for bearings with short or infinitely long length. In this way, there an analytical resolution is possible.

In order to verify the computational model developed in the present work, comparisons will be made with the results obtained by the current literature approximations. According to \cite{Frene}, a bearing can be considered short if the ratio $ L / D $ ($D = 2R_o$ is the diameter) is less than or equal to $ 1/8 $ and long if it is greater than $ 4 $. However, according to \cite{SanAndres:note4}, the model for the short bearing provides accurate results (in the case of simple cylindrical bearings with small to moderate values of the bearing eccentricity, $ e \leq 0.75 F $) already with the ratio $ L / D \leq 1/2 $.

To analyze the short journal bearing, the pressure variation in the $z$ direction is considered much larger than in the $\theta$ direction, i.e. $\partial p/\partial \theta \ll \partial p/\partial z$. Thus, the first term in the Eq. (\ref{u}) is neglected, and one obtains \cite{Yamamoto}:

\begin{equation} \label{eq:pressure_short}
    p = \dfrac{-3\mu \epsilon \omega \sin{\theta}}{\left(R_\theta - R_i\right)^2\left(1 + \epsilon \cos{\theta}\right)^3}\left[\left(z-\dfrac{L}{2}\right)^2 - \dfrac{L^2}{4}\right]\,,
\end{equation}
where $\epsilon = \dfrac{e}{R_o - R_i}$ is the eccentricity ratio and $\omega$ the rotor speed.

\begin{table}[H]
\centering
\begin{tabular}{r|l}
\multicolumn{1}{c|}{\textbf{Element}} & \multicolumn{1}{c}{\textbf{Value}} \\ \hline
Rotor radius $(R_i)$                  & $0.2 \left[m\right]$               \\
Gap $(F)$                  & $0.0001 \left[m\right]$ 
\\
Load $(W)$                   & $50  \left[N\right]$            \\
Ratio $L/D$                      & $1/8 $            \\
Viscosity $(\mu)$                    & $0.015\left[Pa.s\right]$           \\
Rotation speed $(\omega)$                     & $10.472 \left[rad/s\right]$      
\end{tabular}
\caption{Parameters used for the simulation on a short cylindrical bearing.}\label{tab:dados-mancal_curto}
\end{table}

\begin{figure}[h!]
    \centering
    \includegraphics{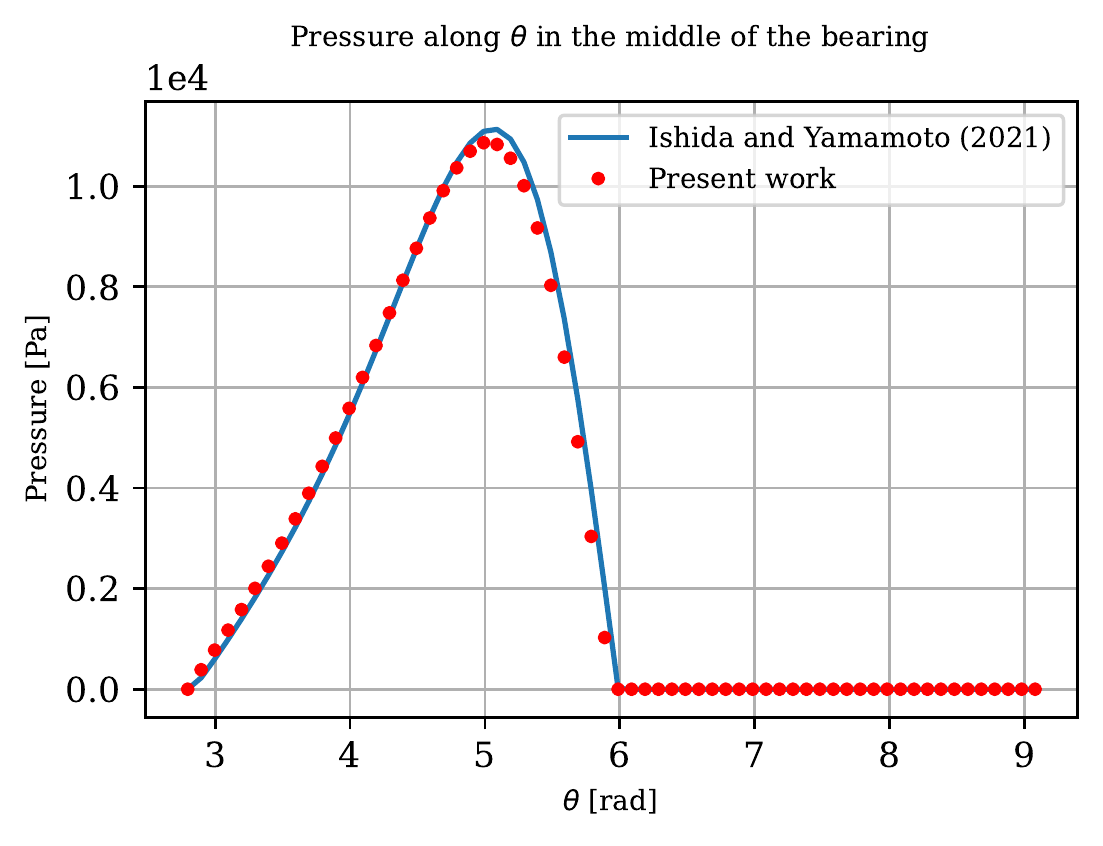}
    \caption{Pressure field obtained for a short journal bearing; Table \ref{tab:dados-mancal_curto}. Comparison of the current results with the one found in \cite{Yamamoto}.} \label{fig: short_pressure}
\end{figure}

The numerical simulation, using data shown in Table \ref{tab:dados-mancal_curto}, generates a relative error $ E \approx 10 ^ {- 3} $. Figure \ref{fig: short_pressure} shows the pressure field computed together with the reference result \cite{Yamamoto}. This result indicates that the present model generate consistent results.

The forces can be obtained integrating the pressure field given in Eq. (\ref{eq:pressure_short}) using Eq. (\ref{eq:force}). The relative errors, between the forces computed numerically and analytically (pressure fields shown in Fig. \ref{fig: short_pressure}), are $ E_N \approx 0.02 $ for the radial force and $ E_T \approx0.005 $ for the tangential force. Concerning the equilibrium position, the are were also small: $ f_x \approx | f_y - W | \approx 10^{-6} [N] $.

The errors in the coefficients are obtained comparing the numerical results with the ones presented in \cite{Friswell}. The following relative errors are obtained:

\begin{align}\label{eq:erros-coeficientes}
    \begin{bmatrix}
    	E_{k_{xx}} & E_{k_{xy}}\\
    	E_{k_{yx}} & E_{k_{yy}}
	\end{bmatrix} \approx
	\begin{bmatrix}
    	0.017 & 0.003\\
    	0.001 & 0.04
	\end{bmatrix}
	\quad \text{e} \quad
	\begin{bmatrix}
    	E_{c_{xx}} & E_{c_{xy}}\\
    	E_{c_{yx}} & E_{c_{yy}}
	\end{bmatrix} \approx
	\begin{bmatrix}
    	0.005 & 0.04\\
    	0.06 & 0.002
	\end{bmatrix}
\end{align}

Now that the proposed computational model was verified with the short bearing approximation found in the literature, the new geometry parameterization, developed in Section \ref{sec:geometria}, can be explored in other bearing configurations, or other machine elements. 



\subsection{Elliptical bearing}

Cylindrical geometry bearings, at high rotation speeds and with small external loads applied, are prone to instability or high vibrations \cite{machado:2011}. This is because, in these situations, the machine tends to operate centered, reducing the variation in the thickness of the lubricating film and, consequently, the ability to generate hydrodynamic pressure from the bearing. Several configurations of bearings were developed in order to circumvent this problem, producing preloads that make the shaft work outside the new established centers, thus creating the desired wedge effect.

\begin{figure}[h!] 
    \centering
    \includegraphics[width=.5\textwidth]{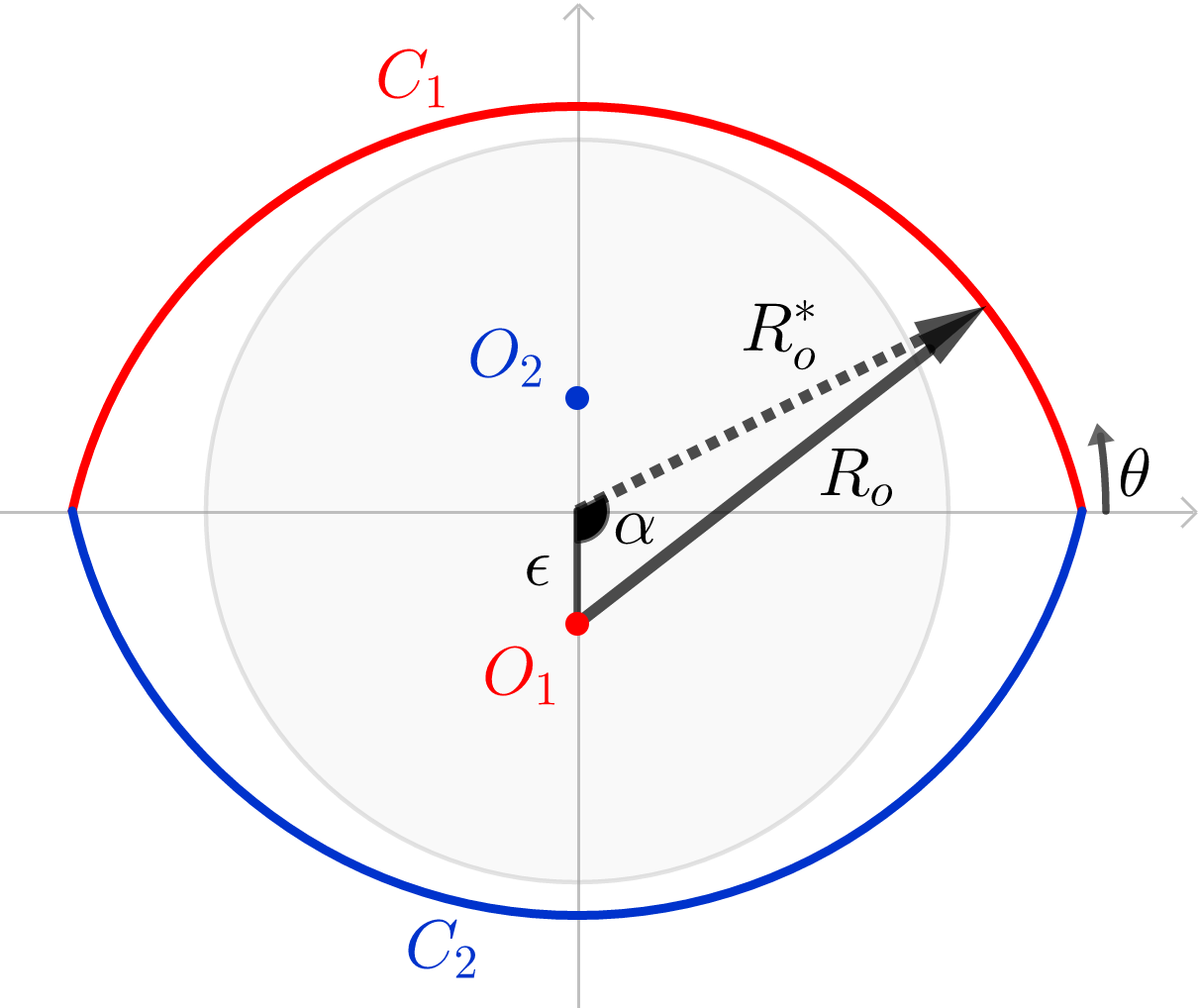}
    \caption{Schematic drawing of an elliptical bearing.} \label{fig:mancaleliptico}
\end{figure}

A feature commonly found in high-speed machines is the elliptical bearing or "lemon bearing", as it is also known \cite{ostayen:2009}. This is a variation of the cylindrical bearing with axial groove and reduced clearance in one direction. Ostayen and Beek \cite{ostayen:2009} claim that this bearing configuration is relatively cheap and easy to manufacture, as it is made from adaptations in the cylindrical bearing itself, being composed of  two circular arcs (cuts from the original structure), with aligned centers, generating two horizontal slits, as shown in Fig. \ref{fig:mancaleliptico}.

To consider this new geometry, stator radius is is no longer constant in $ \theta $. As seen in Fig. \ref{fig:mancaleliptico}, the new stator is composed of the arc $ C_1 $, with center in $ O_1 $, joined to the arc $ C_2 $, centered in $ O_2 $, both with radius $ R_o $. In this new configuration, the centers are at a distance $ \epsilon $ from the origin, called ellipticity. Now, the stator radius is described from the origin. This new distance will be called $ R_o^* $ and it varies along the angular position.

Using the cosine law, the following relation is obtained:

\begin{align}\label{eq:ro*}
    R_o^* = \sqrt{R_o ^2 - \epsilon^2 \sin^2{\alpha}} + \epsilon \cos{\alpha},
\end{align}
where $\alpha =$ 
$\begin{cases} 
\pi/2 + \theta \text{,} 
&\mbox{se} \quad \theta \in \text{1st quadrant} \\ 
3\pi/2 + \theta \text{,} 
&\mbox{se} \quad \theta \in \text{2nd  quadrant} \\
\theta - \pi/2 \text{,} 
&\mbox{se} \quad \theta \in \text{3rd quadrant} \\
5\pi/2 -\theta \text{,} 
&\mbox{se} \quad \theta \in \text{4th quadrant}
\end{cases}.
$\\

Another important parameter to be defined is the preload $ m $ which, in this text, will be established as:
\begin{align}\label{eq:rpreload}
    m = \dfrac{\epsilon}{F}
\end{align}
where $ \epsilon $ is the ellipticity and $ F = R_o - R_i $ is the radial clearance.

For $m=0$, the bearing becomes cylindrical, while when $m$ tends to  $1$ the stator arcs tend to touch the axis. Garner et al \cite{garner:1980} suggest that the maximum preload adopted should be 0.6. According to the authors, values above 0.7 can be used in an effort to improve stability, however, greater lateral clearances are generated, causing the horizontal dynamic coefficients to decrease, in addition to requiring a greater flow of oil.

\begin{table} 
\centering
\begin{tabular}{r|l}
\multicolumn{1}{c|}{\textbf{Element}} & \multicolumn{1}{c}{\textbf{Value}} \\ \hline
Stator radius $(R_o)$                  & $0.015 \left[m\right]$               \\
Gap $(F)$                  & $9 \cdot 10^{-5} \left[m\right]$ 
\\
Length $(L)$                   & $0.02 \left[m\right]$            \\
Load $(W)$                      & $100 \left[N\right]$            \\
Viscosity $(\mu)$                    & $5.449 \cdot 10^{-2}\left[Pa.s\right]$           \\
Rotation speed $(\omega)$                     & $261.8 \left[rad/s\right]$
\end{tabular}
\caption{Numerical simulation for the elliptical bearing.}\label{tab:mancal-eliptico}
\caption*{\textbf{Fonte:} Adapted from \citep{machado:2006}}
\end{table}

\begin{figure}
    \centering
    \begin{subfigure}[b]{0.49\textwidth}
        \centering
        \includegraphics{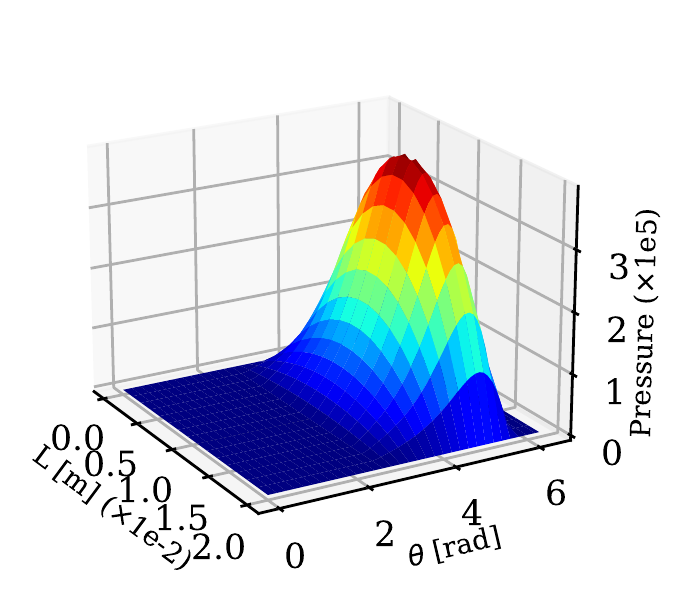}
        \caption{Pressure distribution.}
    \end{subfigure}
    \begin{subfigure}[b]{0.49\textwidth}
        \centering
        \includegraphics{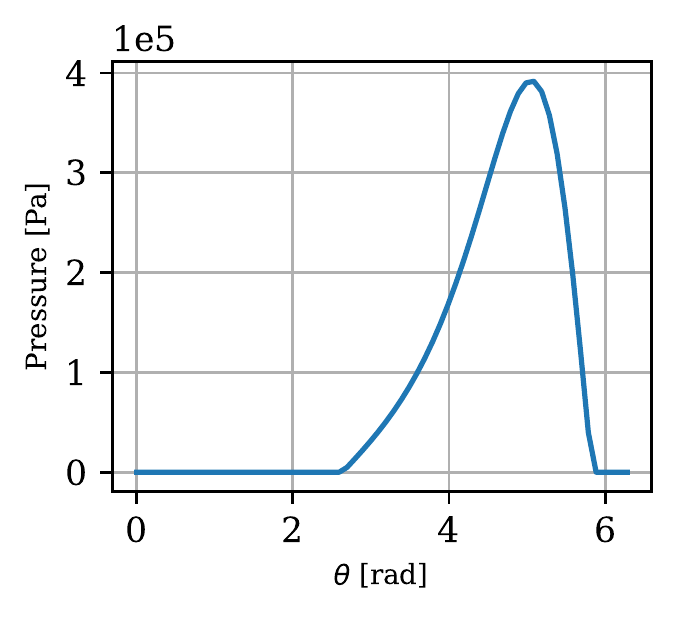}
        \caption{Side view.}
    \end{subfigure}
\caption{Cylindrical bearing $(m=0)$.}\label{fig:pressao-elip-m0}
\end{figure}

\begin{figure}
    \centering
    \begin{subfigure}[b]{0.49\textwidth}
        \centering
        \includegraphics{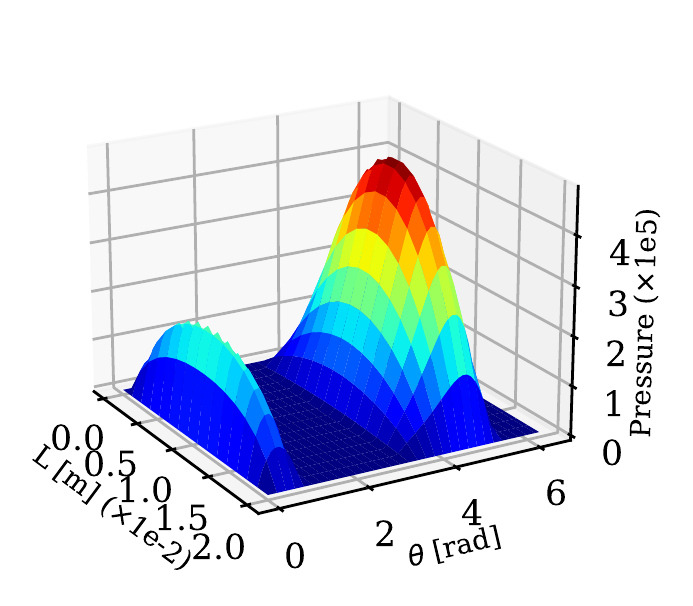}
        \caption{Pressure distribution.}
    \end{subfigure}
    \begin{subfigure}[b]{0.49\textwidth}
        \centering
        \includegraphics{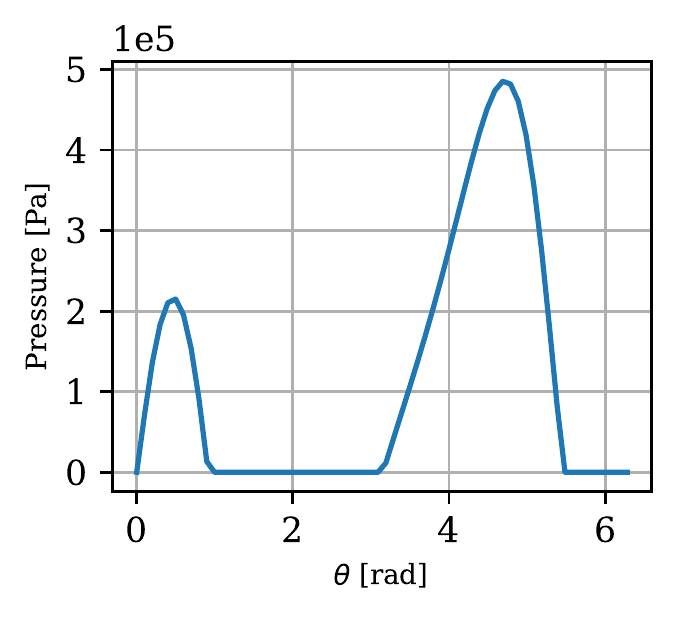}
        \caption{Side view.}
    \end{subfigure}
\caption{Elliptical bearing with $(m=0.4)$.}\label{fig:pressao-elip-m04}
\end{figure}

\begin{figure}
    \centering
    \begin{subfigure}[b]{0.49\textwidth}
        \centering
        \includegraphics{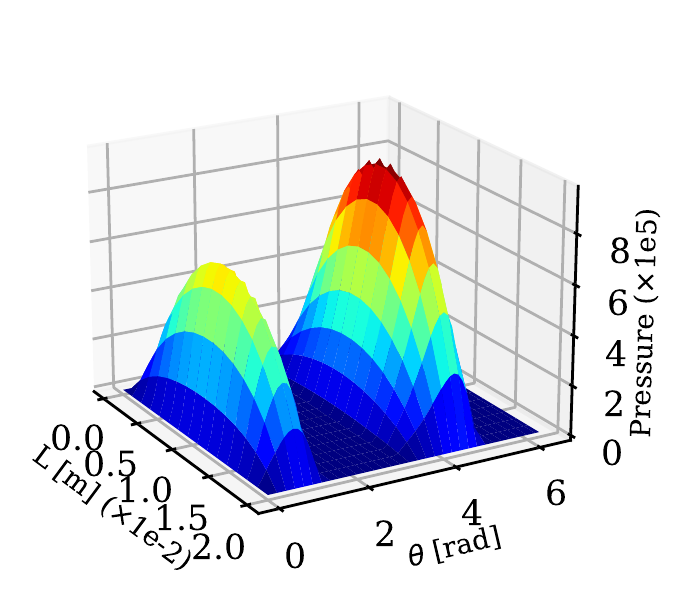}
        \caption{Pressure distribution.}
    \end{subfigure}
    \begin{subfigure}[b]{0.49\textwidth}
        \centering
        \includegraphics{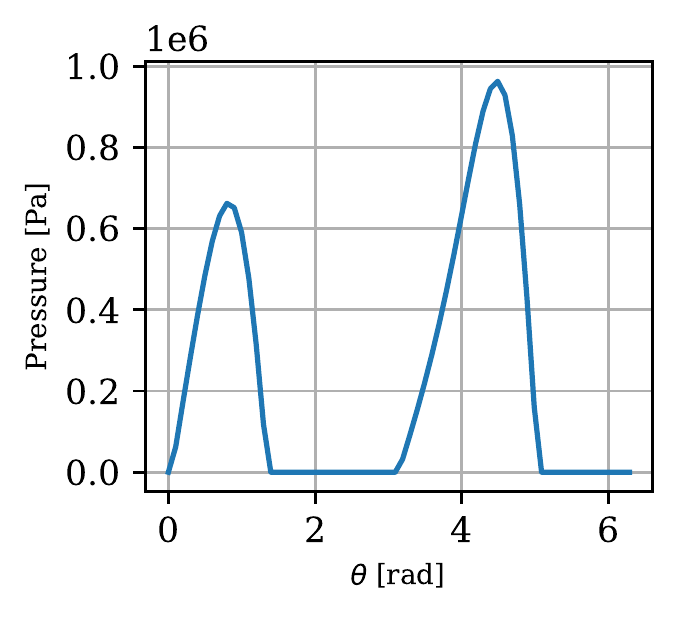}
        \caption{Side view.}
    \end{subfigure}
\caption{Elliptical bearing with $(m=0.6)$.}\label{fig:pressao-elip-m06}
\end{figure}

Given the data shown in Table \ref{tab:mancal-eliptico}, the results for the pressure on the elliptical bearing, using the modeling presented in this work, are compatible with those exposed in \cite{machado:2006}. Figures \ref{fig:pressao-elip-m0} (a-b) illustrate the pressure field for a bearing described as elliptical, but with $ m = 0 $, that is, cylindrical. In contrast, the graphs shown in Figs. \ref{fig:pressao-elip-m04} (a-b) and \ref{fig:pressao-elip-m06} (a-b), indicate the pressure behavior for preloads equal to $ 0.4 $ and $ 0.6 $, respectively.

It can be seen that Figs. \ref{fig:pressao-elip-m04} and \ref{fig:pressao-elip-m06} show two pressure peaks: one for the upper bearing arch ($0$ to $\pi$) and one for the lower arc ($ \pi $ to $ 2\pi $). This phenomenon can be explained due to the configuration of the elliptical bearing. Even if the shaft operates centered, there will be variations in the thickness of the fluid, causing the oil film to be compressed, both in the upper and lower arc.

\begin{figure}[H] 
    \centering
    \includegraphics{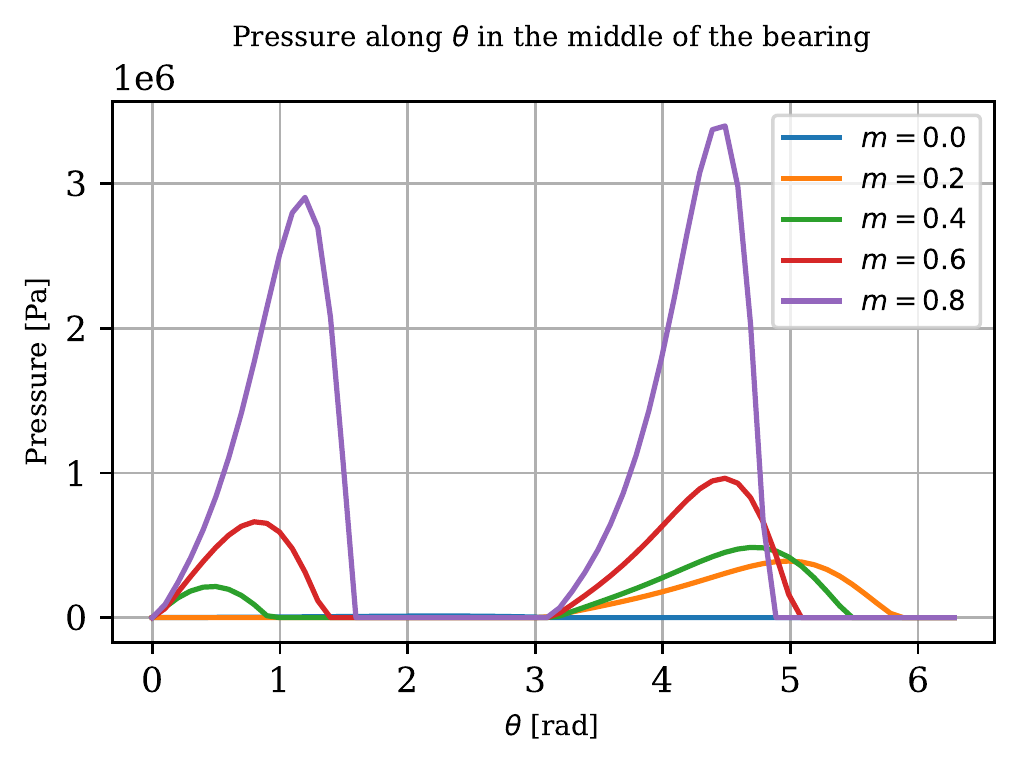}
    \caption{Pressure along $ \theta $ in the middle of the elliptical bearing for different values of the preload $ m $.} \label{fig:mancaleliptico-comparacao}
\end{figure}

Analyzing also the graph of Fig. \ref{fig:mancaleliptico-comparacao}, it is understood that the reaction force increases when the preload increases, and, hence the pressure. For $ m = 0.8 $, the pressures generated are considerably higher than for other preloads, being an option for cases in which a greater demand is sought to rotor stability.

\begin{figure}
    \centering
    \begin{subfigure}[h!]{0.49\textwidth}
        \centering
        \includegraphics{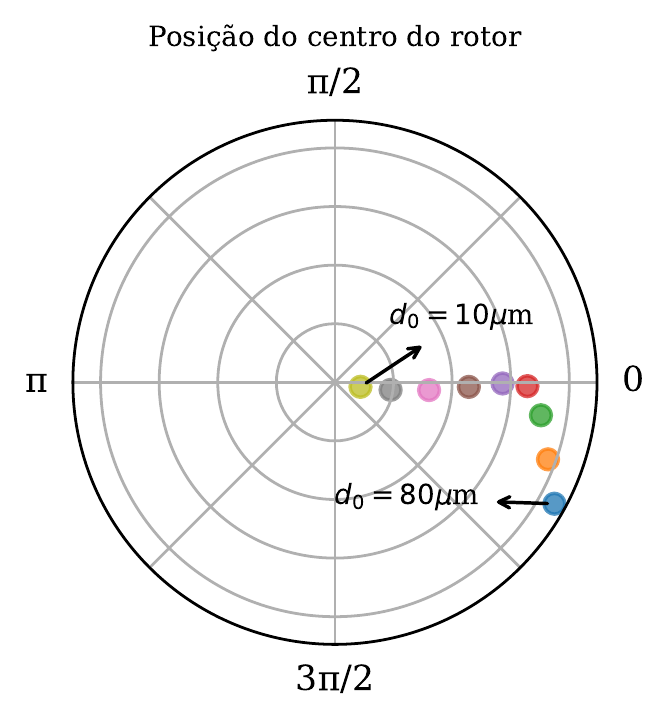}
        \caption{Rotor center.}\label{fig:elip_centrorotor_x_m}
    \end{subfigure}
    \begin{subfigure}[h!]{0.49\textwidth}
        \centering
        \includegraphics{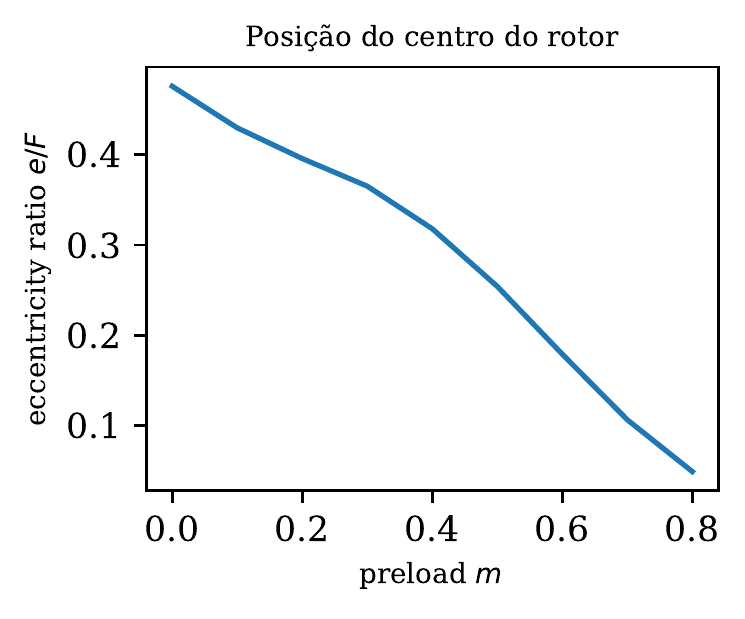}
        \caption{Variation of eccentricity.}\label{fig:elip_excentricidade_x_m}
    \end{subfigure}
\caption{Variation of the equilibrium position on the elliptical bearing for different preloads.}
\end{figure}

With regard to the equilibrium position, Fig. \ref{fig:elip_centrorotor_x_m} illustrates a different behavior comparing to the cylindrical bearing. That is, the Sommerfeld Number used in the cylindrical bearing is not valid to analyze the static performance characteristics of the elliptical bearing. For the elliptical bearing, increasing the preload, first the rotor moves vertically and, after $ m = 0.3 $, it walks in the horizontal direction, approaching the center of the stator, always reducing the eccentricity (Fig. \ref{fig:elip_excentricidade_x_m}).

\begin{figure}
    \centering
    \includegraphics{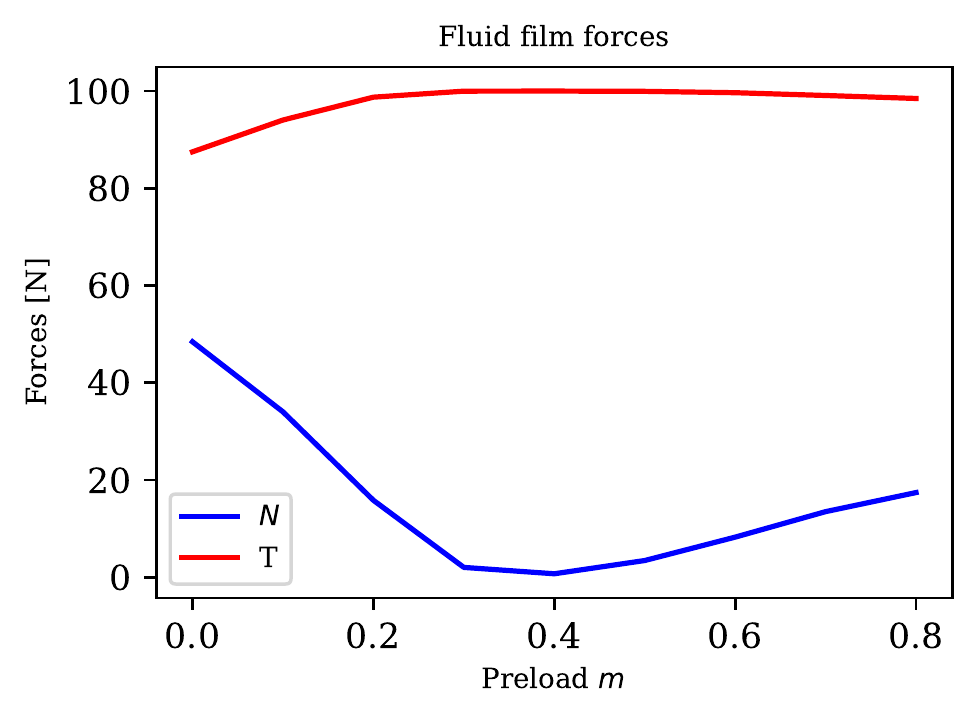}
    \caption{Behavior of radial and tangential forces on the elliptical bearing for different preloads.}
    \label{fig:elip_forcas_x_m}
\end{figure}

The forces of the fluid film can be seen in Fig. \ref{fig:elip_forcas_x_m}, in which it is evident that the tangential force is always much higher than the radial force for any preload applied to the elliptical bearing. 

\begin{figure}
    \centering
    \begin{subfigure}[b]{0.49\textwidth}
        \centering
        \includegraphics{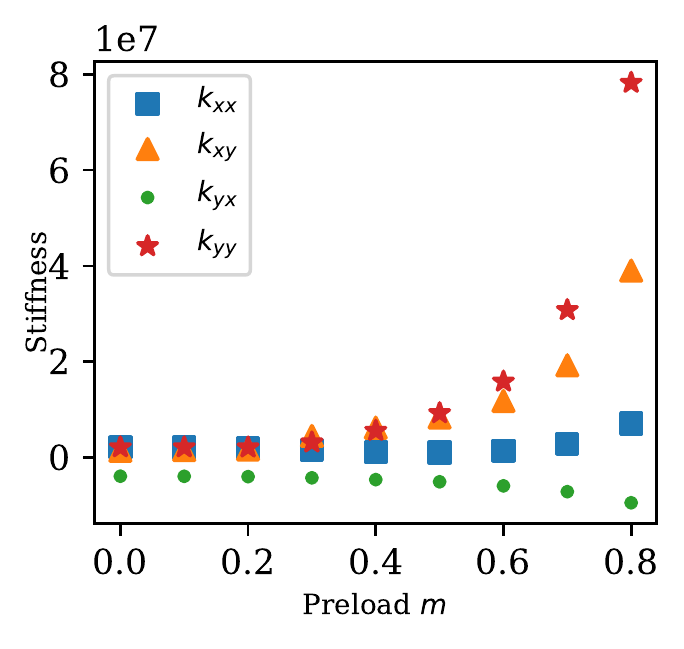}
        \caption{Stiffness coefficients.}
    \end{subfigure}
    \begin{subfigure}[b]{0.49\textwidth}
        \centering
        \includegraphics{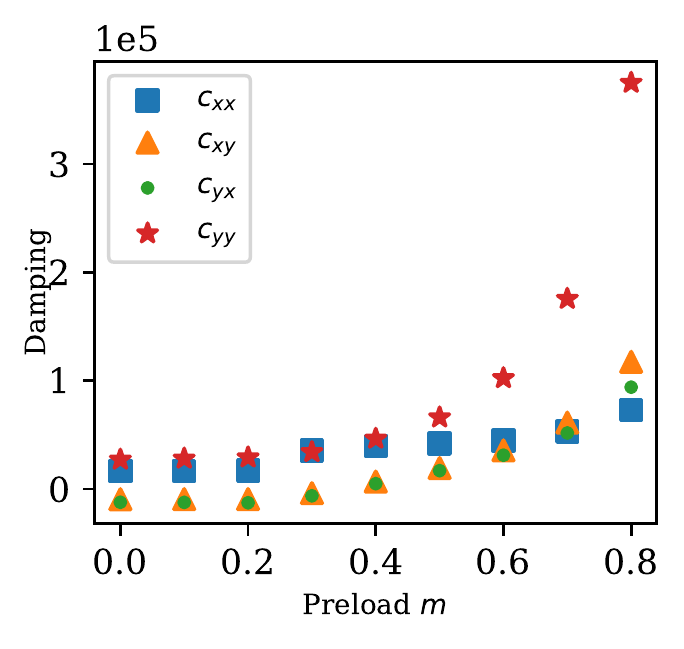}
        \caption{Damping coefficients.}
    \end{subfigure}
\caption{Dynamic coefficients of the elliptical bearing for different preloads.}\label{fig:elip_coeficientes_x_m}
\end{figure}

It is also noticeable, when analyzing Figs. \ref{fig:elip_coeficientes_x_m} (a-b), that the dynamic coefficients of the elliptical bearing suffer great variation between their terms. For the preload $ m $ up to $ 0.2 $, the terms $ k_{xx} $, $ k_{xy} $ and $ k_{yy} $ are very close to each other. From $ m = 0.3 $ it is observed that the difference between these terms starts to increase. The cross term $ k_{xy} $ is increasingly larger than the direct term $ k_{xx} $, which might indicate instability, as pointed out by \cite{machado:2006}. Concerning the damping coefficients, all terms increase as the preload increases. It is worth remembering that the main function of damping is to mitigate the effects of vibration on the system, and that high values of these coefficients are desired.

The difference between the results of the cylindrical and elliptical bearing is clear. In the first configuration, all the factors that reduce the eccentricity imply a decrease in the pressure field, forces and variation of the dynamic coefficients. For the elliptical bearing configuration, the exact opposite occurs. By increasing the preload, rotor and stator become closer to concentricity and, this yields an increase in all the results just mentioned. However, the increase of the stiffness cross coefficient indicates that the system risks to be instable.

\subsection{Wear bearing}

Although lubrication reduces the friction between the metallic surfaces of the bearing, these structures usually suffer wear after a long operating period or due to a certain number of starting cycles. One of the most common causes of interruptions in rotating systems is the occurrence of failures related to hydrodynamic bearings \cite{machado:2015}. As the wear suffered by these components affects the radial clearance, they certainly influence the results of the pressure field and, consequently, the reaction forces and dynamic coefficients. For this reason, it is relevant to investigate the effect of wear, under certain operating conditions, on the dynamic response of the system.

The wear geometry that will be used in this text was adapted from  \cite{machado:2015}. The description was initially proposed by \cite{dufrane:1983} and validated experimentally  \cite{hashimoto:2008}.

\begin{figure} 
    \centering
    \includegraphics[width=.5\textwidth]{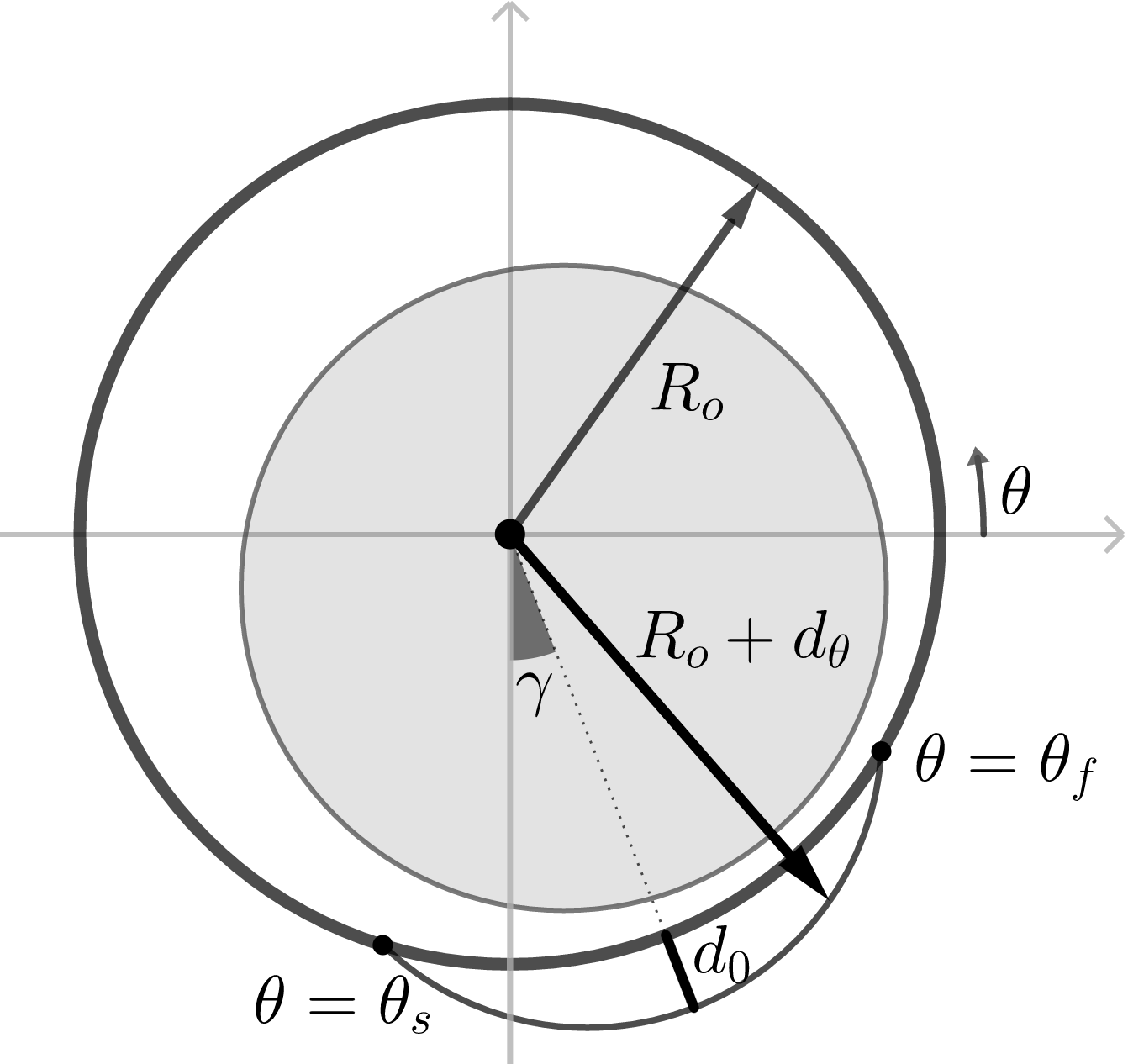}
    \caption{Schematic drawing of a worn bearing.} \label{fig:mancaldesgaste}
\end{figure}

Dufrane et al \cite{dufrane:1983} consider the wear thickness uniform in the axial direction, and located at the bottom of the bearing, symmetrically around the $ Y $ axis. On the other hand, Machado and Cavalca \cite{machado:2015} states that, for practical cases, most wear and tear occurs in the fourth quadrant, the location of the axis balance position for an anti-clockwise rotation. For this reason, taking into account that the spent region is symmetrical in relation to the greatest depth $ d_0 $, an angular displacement $ \gamma $ will be considered from the vertical axis to the deepest part, as shown in Fig. \ref{fig:mancaldesgaste}.

To include wear in the geometry, it is necessary to make some adaptations to the stator radius. Considering that the fault starts at the angular position $ \theta = \theta_s $ and ends at $ \theta = \theta_f $, the stator description from the origin is defined as

\begin{align}\label{ro*-wear}
    R_o^* = R_o + d_{\theta}
\end{align}
where $d_{\theta} =$
$\begin{cases} 
    0 \text{,} 
    &\mbox{se} 
    \quad 0 \leq \theta \leq \theta_s\text{,} \quad
    \theta_f \leq \theta \leq 2\pi \\ 
    d_0 - F \left(1 + \cos{\left(\theta - \pi/2\right)} \right) \text{,} 
    &\mbox{se} 
    \quad \theta_s < \theta < \theta_f
\end{cases}$.

In $ \theta_s $ and $ \theta_f $, the wear depth is zero, so the location of the edges can be defined as follows:

\begin{align}
    \theta_s = \pi/2 + \cos^{-1}{\left(d_0/F -1\right)} + \gamma \nonumber\\
    \theta_f = \pi/2 - \cos^{-1}{\left(d_0/F -1\right)} + \gamma 
\end{align}

\begin{table} 
\centering
\begin{tabular}{r|l}
\multicolumn{1}{c|}{\textbf{Element}} & \multicolumn{1}{c}{\textbf{Value}} \\ \hline
Stator radius $(R_o)$                  & $0.015 \left[m\right]$               \\
Gap $(F)$                  & $9 \cdot 10^{-6} \left[m\right]$ 
\\
Length $(L)$                   & $0.02 \left[m\right]$            \\
Load $(W)$                      & $18.9 \left[N\right]$            \\
Viscosity $(\mu)$                    & $0.1044\left[Pa.s\right]$           \\
Rotation speed $(\omega)$                     & $104.72 \left[rad/s\right]$
\end{tabular}
\caption{Numerical simulation for the worn bearing.}\label{tab:mancal-desgaste}
\caption*{\textbf{Fonte:} Adapted from \citep{machado:2015}}
\end{table}

\begin{figure}
    \centering
    \begin{subfigure}[b]{0.49\textwidth}
        \centering
        \includegraphics{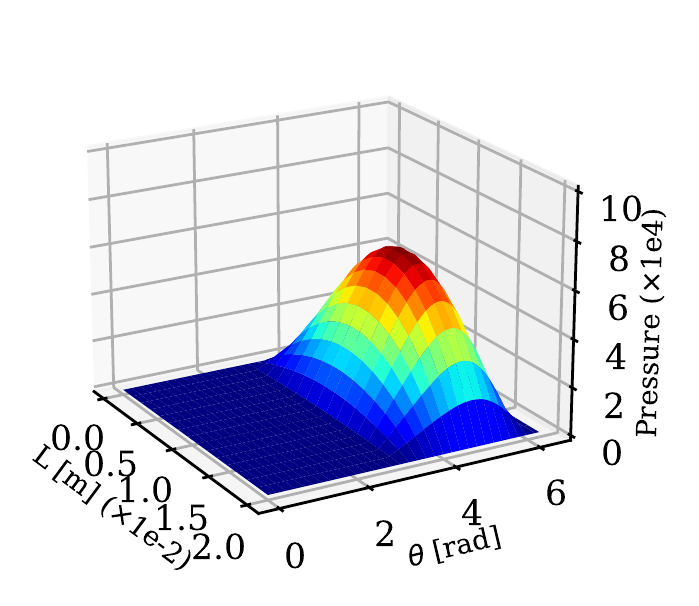}
        \caption{Pressure distribution.}
    \end{subfigure}
    \begin{subfigure}[b]{0.49\textwidth}
        \centering
        \includegraphics{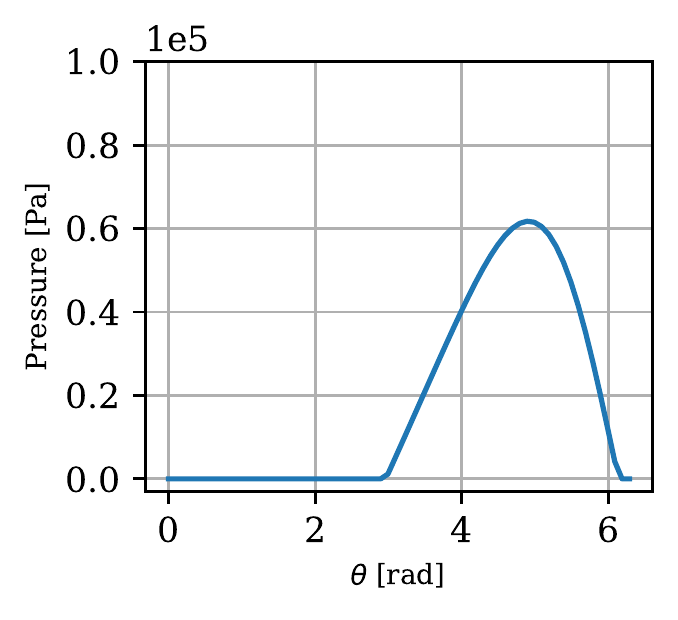}
        \caption{Side view.}
    \end{subfigure}
\caption{Bearing without wear.}\label{fig:pressao-wear-d0_gama0}
\end{figure}

\begin{figure}
    \centering
    \begin{subfigure}[b]{0.49\textwidth}
        \centering
        \includegraphics{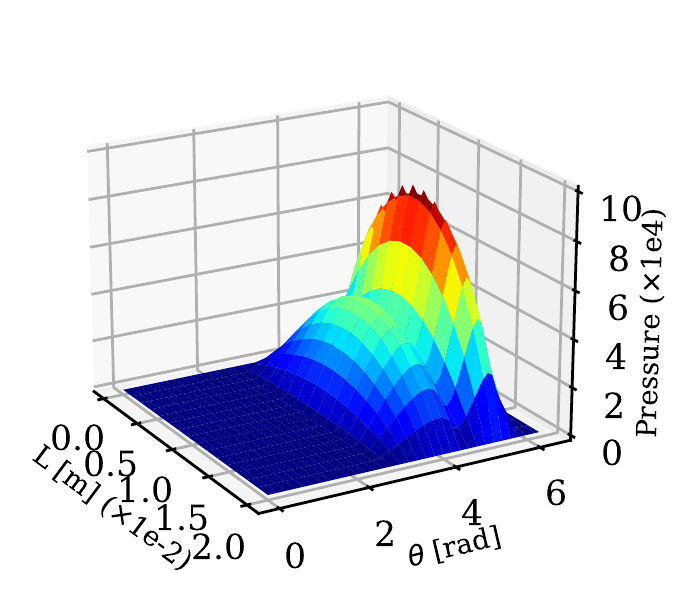}
        \caption{Pressure distribution.}
    \end{subfigure}
    \begin{subfigure}[b]{0.49\textwidth}
        \centering
        \includegraphics{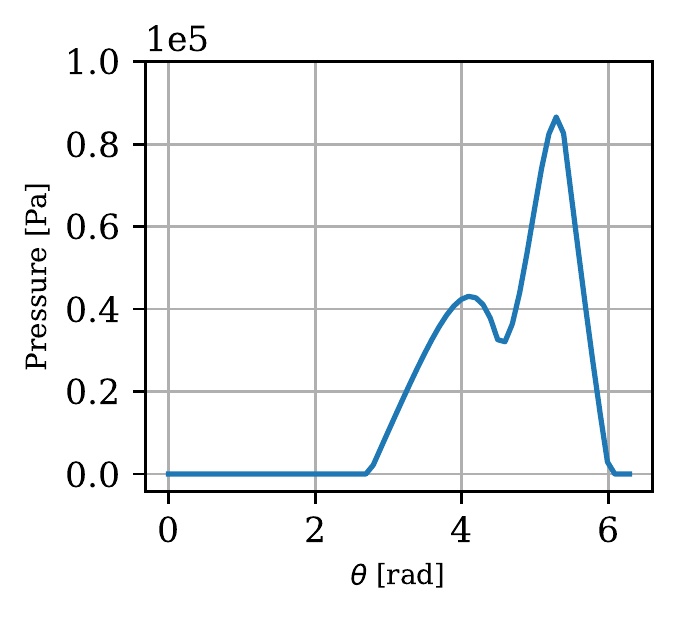}
        \caption{Side view.}
    \end{subfigure}
\caption{Wear bearing ($d_0 = 10[\mu m]$, $\gamma = 10^{\circ}$).}\label{fig:pressao-wear-d10_gama10}
\end{figure}

\begin{figure}
    \centering
    \begin{subfigure}[b]{0.49\textwidth}
        \centering
        \includegraphics{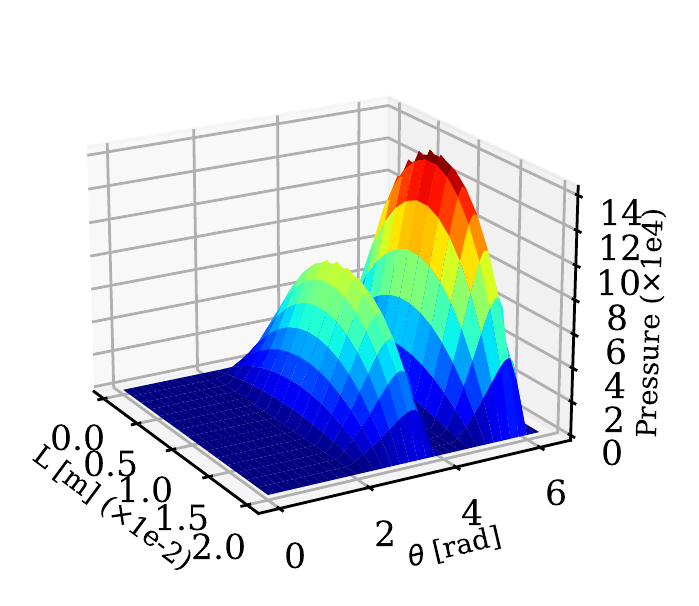}
        \caption{Pressure distribution.}
    \end{subfigure}
    \begin{subfigure}[b]{0.49\textwidth}
        \centering
        \includegraphics{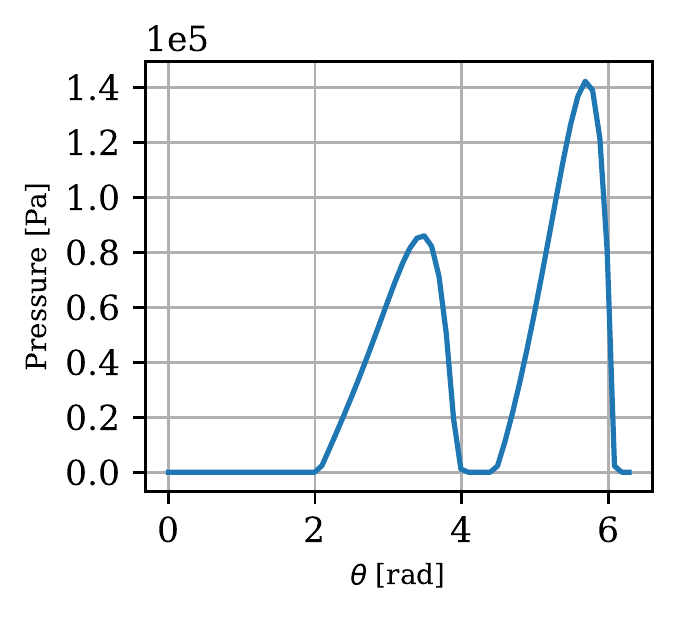}
        \caption{Side view.}
    \end{subfigure}
\caption{Wear bearing ($d_0 = 50[\mu m]$, $\gamma = 10^{\circ}$).}\label{fig:pressao-wear-d50_gama10}
\end{figure}

\begin{figure}
    \centering
    \begin{subfigure}[b]{0.49\textwidth}
        \centering
        \includegraphics{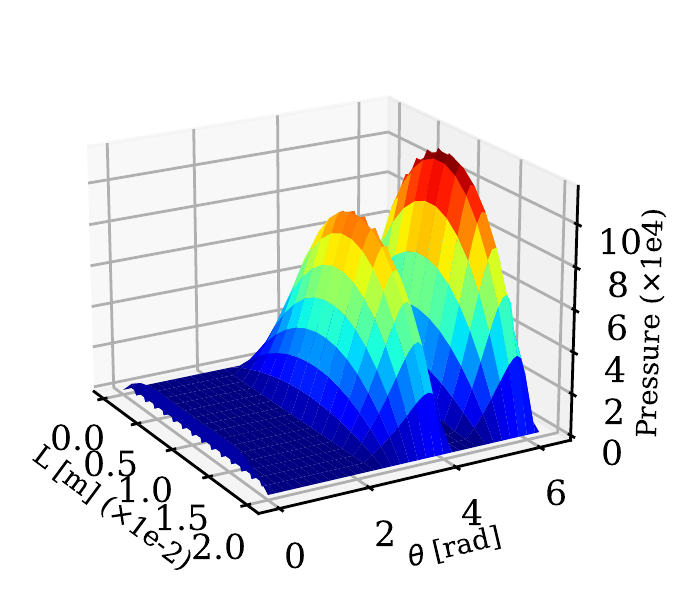}
        \caption{Pressure distribution.}
    \end{subfigure}
    \begin{subfigure}[b]{0.49\textwidth}
        \centering
        \includegraphics{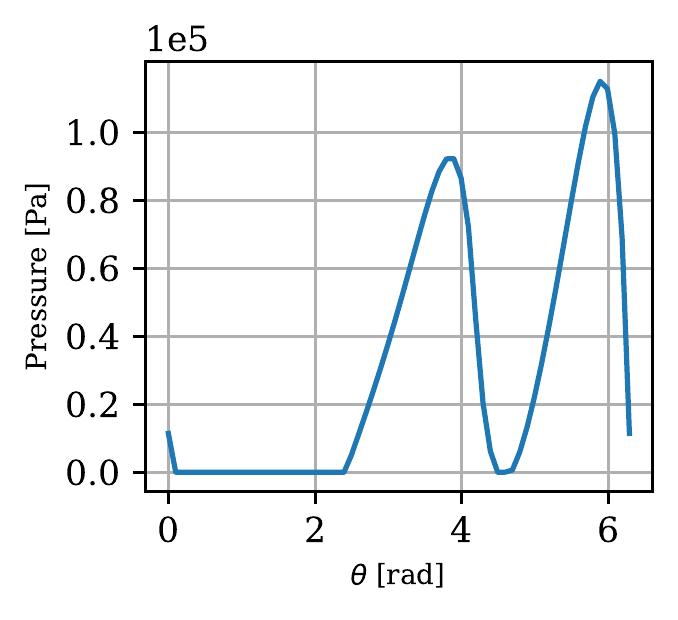}
        \caption{Side view.}
    \end{subfigure}
\caption{Wear bearing ($d_0 = 50[\mu m]$, $\gamma = 30^{\circ}$).}\label{fig:pressao-wear-d50_gama30}
\end{figure}

With the geometry described as above, and data shown in Table \ref{tab:mancal-desgaste}, the results for the pressure field in the bearing with wear are verified  with \cite{machado:2015}. For a cylindrical bearing without wear, the pressure field is represented by Figs. \ref{fig:pressao-wear-d0_gama0} (a-b). Figs. \ref{fig:pressao-wear-d10_gama10} (a-b) illustrate the pressure field for the bearing with maximum depth wear $ d_0 = 10 [\mu m] $ and an offset $ \gamma = 10^{\circ}\approx 0.17 [rad] $. In Figs. \ref{fig:pressao-wear-d50_gama10} (a-b) $ d_0 = 50 [\mu m] $ is considered for the same displacement and, finally, Figs. \ref{fig:pressao-wear-d50_gama30} (a-b) show the pressure for the same wear with displacement $ \gamma = 30^{\circ} \approx 0.52 [rad] $.


\begin{figure}[H] 
    \centering
    \includegraphics{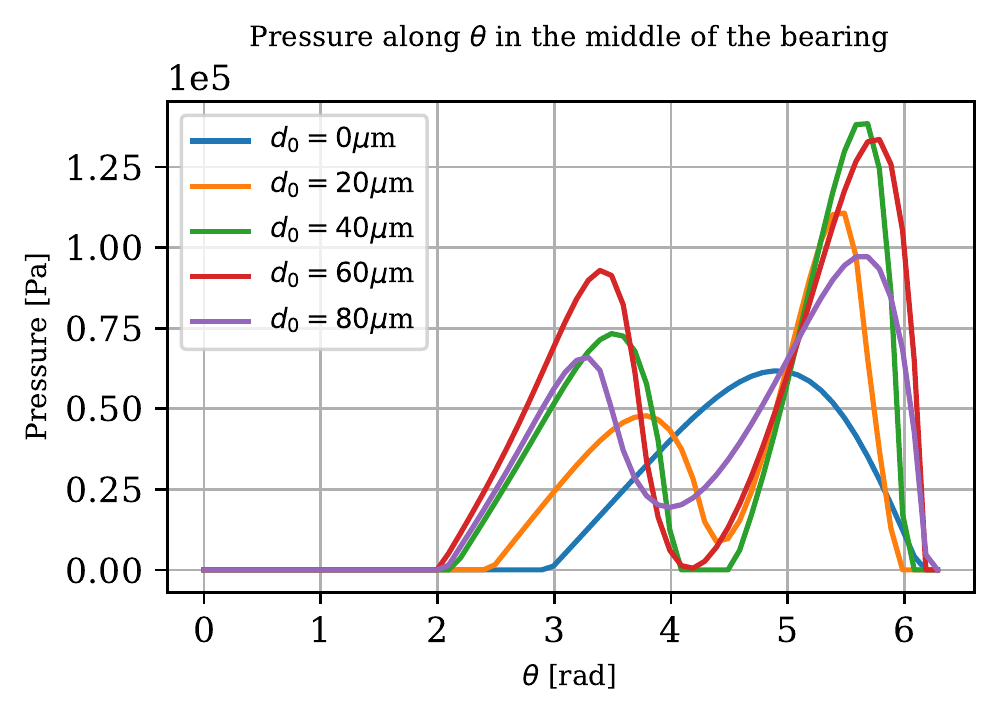}
    \caption{Pressure along $ \theta $ in the middle of the bearing with wear to different depths $d_0$ ($\gamma = 10^{\circ}$).} \label{fig:wear_pressao_x_d}
\end{figure}

It is possible to see that, even for minor wears, the pressure field is affected drastically. In addition to presenting two peaks, as occurred in the elliptical bearing, the magnitude of the pressure is also changed. Fig. \ref{fig:wear_pressao_x_d} shows that, as the depth $ d_0 $ increases, the shape of the pressure field changes considerably.

\begin{figure}
    \centering
    \begin{subfigure}[b]{0.49\textwidth}
        \centering
        \includegraphics{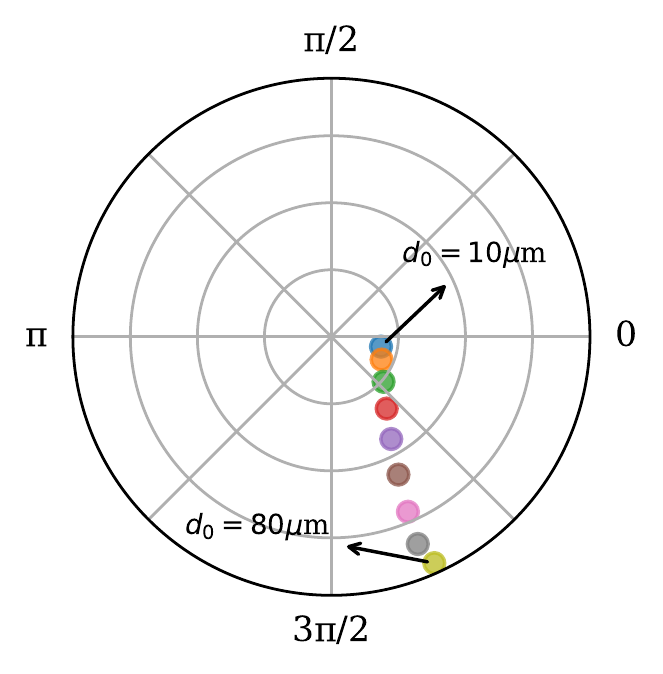}
        \caption{Rotor center.}
    \end{subfigure}
    \begin{subfigure}[b]{0.49\textwidth}
        \centering
        \includegraphics{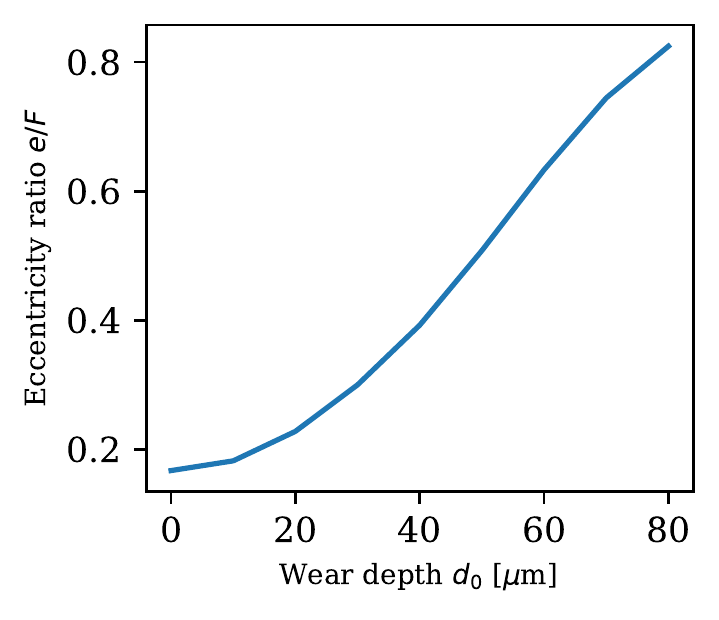}
        \caption{Variation of eccentricity.}
    \end{subfigure}
\caption{Variation of the balance position in the bearing with wear for different depths $d_0$ ($\gamma = 10^{\circ}$).}\label{fig:wear_equilibrio_x_d}
\end{figure}

The depth of wear on the structure also influences the equilibrium position of the rotor, as seen in Figs. \ref{fig:wear_equilibrio_x_d}. As the depth $ d_0 $ increases, the center of the rotor moves away from the center of the stator, increasing the eccentricity. In addition to the direction opposite to the center of the stator, the path taken also differs in behavior when compared to the cylindrical bearing without wear. As the wear depth increases, the center of the rotor moves further down, describing a course close to a straight line.

\begin{figure}
    \centering
    \includegraphics{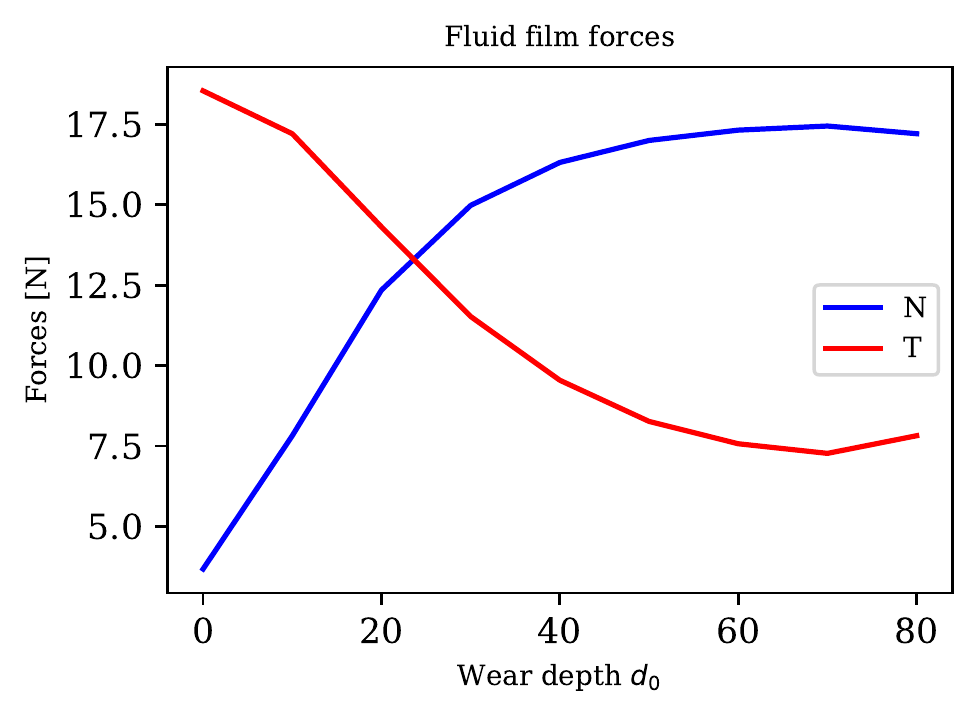}
    \caption{Behavior of radial and tangential forces on the bearing with wear to different depths $d_0$ ($\gamma = 10^{\circ}$).}
    \label{fig:wearforcasxd}
\end{figure}

The radial and tangential forces for the worn bearing develop a different behavior, comparing to the elliptical bearing. Figure \ref{fig:wearforcasxd} indicates that the radial component of the force increases rapidly at greater depths of wear, while the value of the tangential component decreases.

\begin{figure}
    \centering
    \begin{subfigure}[b]{0.49\textwidth}
        \centering
        \includegraphics{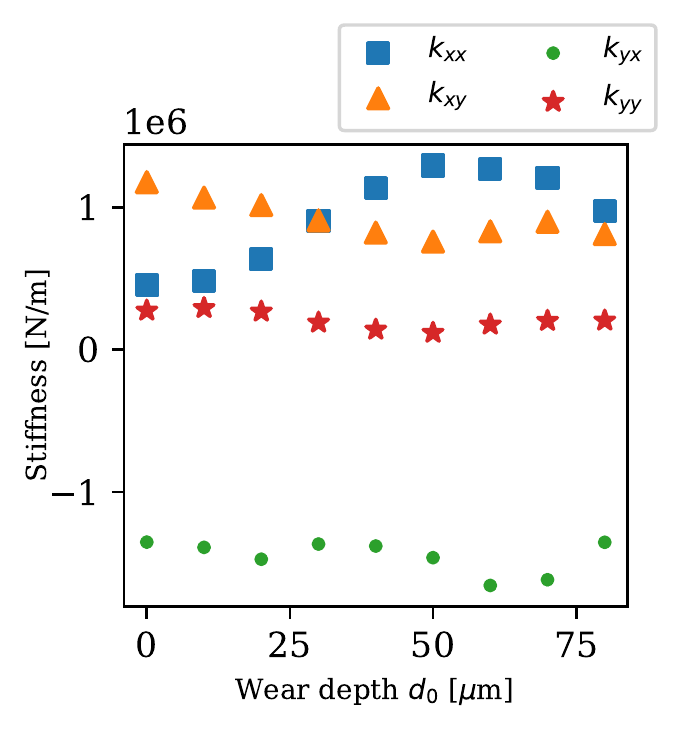}
        \caption{Stiffness coefficients.}
    \end{subfigure}
    \begin{subfigure}[b]{0.49\textwidth}
        \centering
        \includegraphics{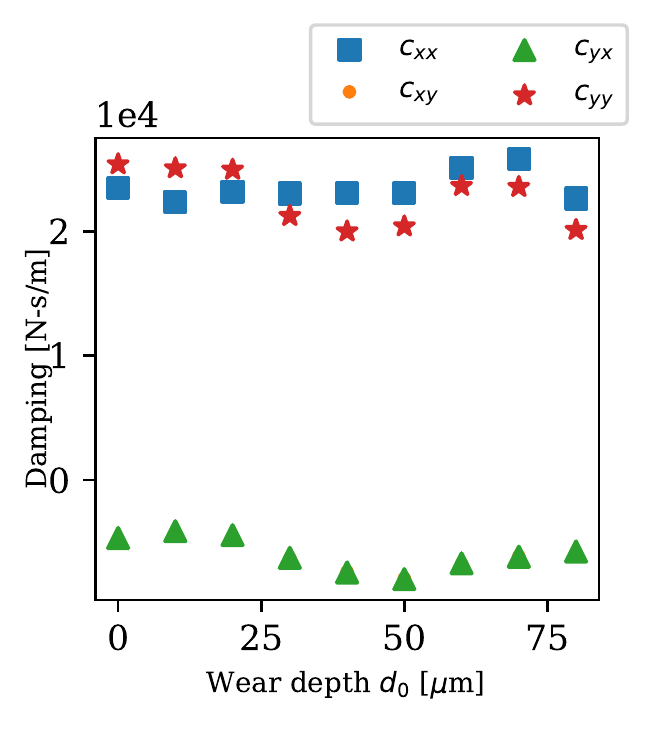}
        \caption{Damping coefficients.}
    \end{subfigure}
\caption{Dynamic bearing coefficients with wear for different depths $d_0$ ($\gamma = 10^{\circ}$).}\label{fig:wear_coeficientes_x_d}
\end{figure}

Finally, with regard to the dynamic coefficients, Figs. \ref{fig:wear_coeficientes_x_d} (a-b) describe the behavior of stiffness and damping for different wear depths. There is an oscillation in the stiffness coefficients, with all the terms distinct from each other and only negative $ k_{yx} $ values. For maximum depth wear up to $ d_0 = 30 [\mu m] $, the cross term $ k_{xy} $ is greater than the direct term $ k_{xx} $, indicating possibility of instability. From that value, $ k_{xx} $ grows, distancing itself from the cross term, but then decreases again shortly thereafter. The coefficients related to the vertical forces do not show very relevant variations, the direct term being always much higher than the cross term. The damping coefficients develop positive values in their direct terms, and negative and similar values in their crossed coefficients.

Thus, it can be concluded that wear on the cylindrical bearing are variations in geometry that directly affect the dynamic response of the system. The stiffness coefficients show intervals in which the risk of instabilities is more critical, which can serve to indicate operational conditions that should be avoided.

\subsection{Comparison between geometries}

Nowadays, many rotating machines operate at high rotation speeds. This speed influences the entire dynamic response of the bearing, from the equilibrium position to the stiffness and damping coefficients. For this reason, it is essential to understand how the model's responses behave for different rotation speeds.

\begin{figure} 
    \centering
    \includegraphics[width=.5\textwidth]{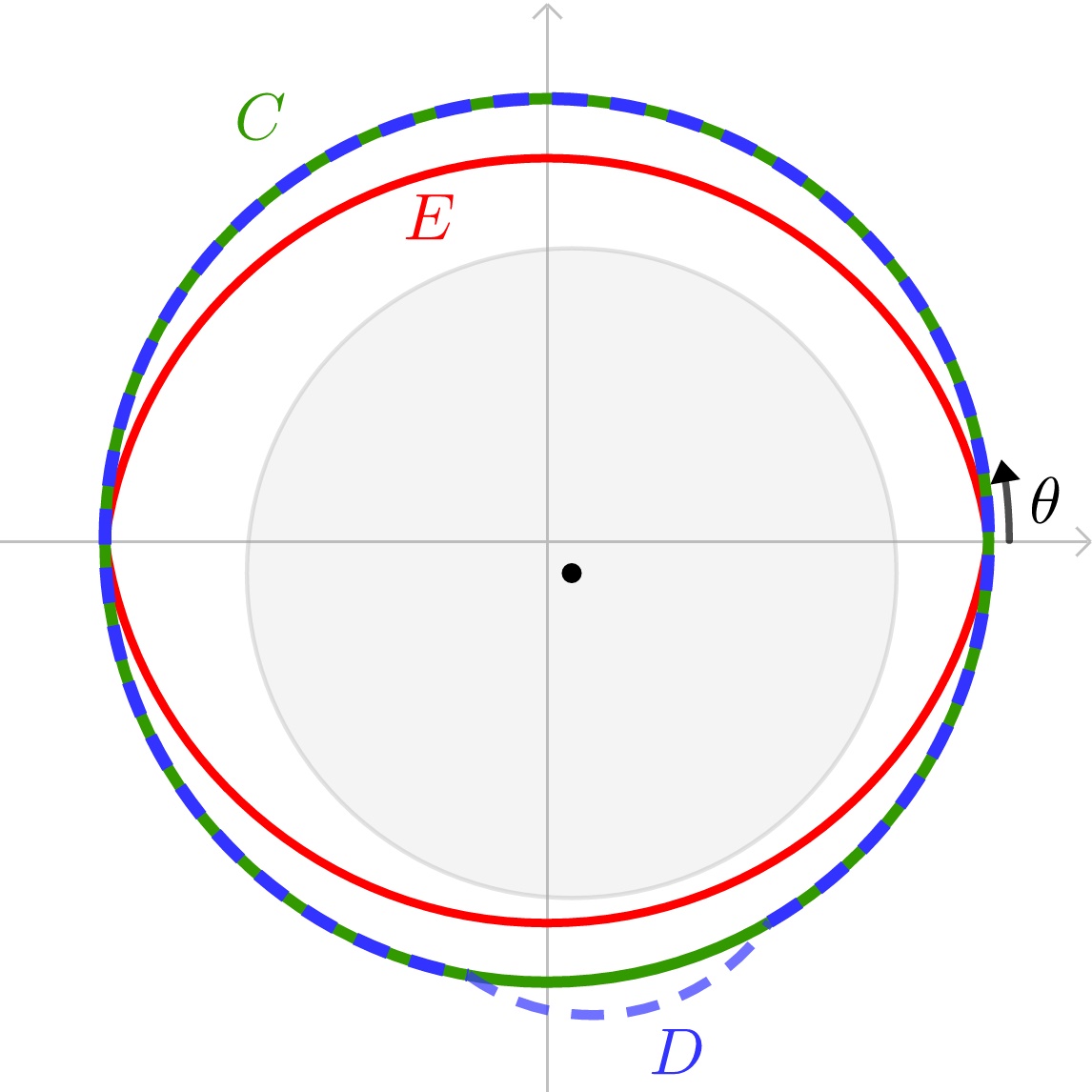}
    \caption{Schematic drawing listing the three bearing configurations shown.} \label{fig:comparacoes}
\end{figure}

\begin{table}
\centering
\begin{tabular}{r|l}
\multicolumn{1}{c|}{\textbf{Element}} & \multicolumn{1}{c}{\textbf{Value}} \\ \hline
Stator radius $(R_o)$                  & $0.015 \left[m\right]$               \\
Gap $(F)$                  & $9 \cdot 10^{-5} \left[m\right]$ 
\\
Length $(L)$                   & $0.02 \left[m\right]$            \\
Load $(W)$                      & $100 \left[N\right]$            \\
Viscosity $(\mu)$                    & $0.05449\left[Pa.s\right]$           \\
Rotation speed $(\omega)$                    &  $50 \leq \omega \leq 550 \left[rad/s\right]$           \\
\end{tabular}
\caption{Numerical simulation for the comparison of bearing geometries.}\label{tab:mancal-comparacao}
\end{table}

In order to compare the results, the data shown in Table \ref{tab:mancal-comparacao} is used, adapting only the specifics of each geometry and varying the rotation speed in the range of $50 \leq \omega \leq 550 \left[rad/s\right]$. For the elliptical bearing the preload $ m = 0.4 $ was considered and, for the bearing with wear, the maximum depth $ d_0 = 50 [\mu m] $ with an angular displacement $ \gamma = 10^{\circ } $. The schematic drawing that relates the three bearing configurations is illustrated in Fig. \ref{fig:comparacoes}, in which the geometries are indicated by the letters $ C $, $ E $ and $ D $, representing cylindrical, elliptical and worn bearings, respectively.

\begin{figure}
    \centering
    \begin{subfigure}[b]{0.49\textwidth}
        \centering
        \includegraphics{ 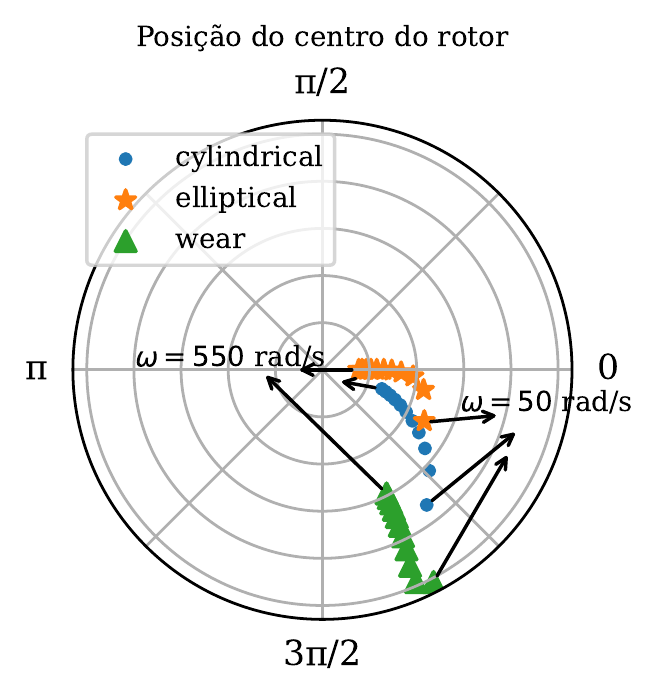}
        \caption{Rotor center.}
    \end{subfigure}
    \begin{subfigure}[b]{0.49\textwidth}
        \centering
        \includegraphics{ 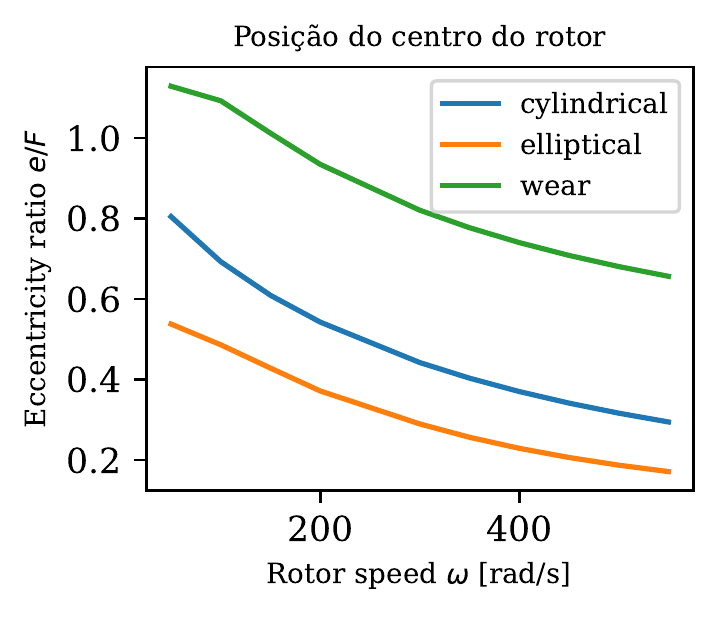}
        \caption{Variation of eccentricity.}
    \end{subfigure}
\caption{Balance position for different bearing configurations, varying the rotation speed.}\label{fig:equilibrio_cil_elip_wear}
\end{figure}

Figures. \ref{fig:equilibrio_cil_elip_wear} (a-b) illustrate the variation of the equilibrium position for the cylindrical, elliptical and worn bearing. In the three cases, the eccentricity decreases with the increase of the rotation speed. However, it is possible to observe the different paths. While the cylindrical bearing moves in a semicircle-like path, the elliptical bearing generates a steeper curve, in which most points are aligned with the horizontal axis. On the other hand, the worn bearing has much higher eccentricities for all speeds. It is interesting to highlight the difference between the bearing balance positions with and without wear. Despite having very similar geometry, wear alters the position of the rotor center and the way it behaves at different speeds.

\begin{figure} 
    \centering
    \includegraphics{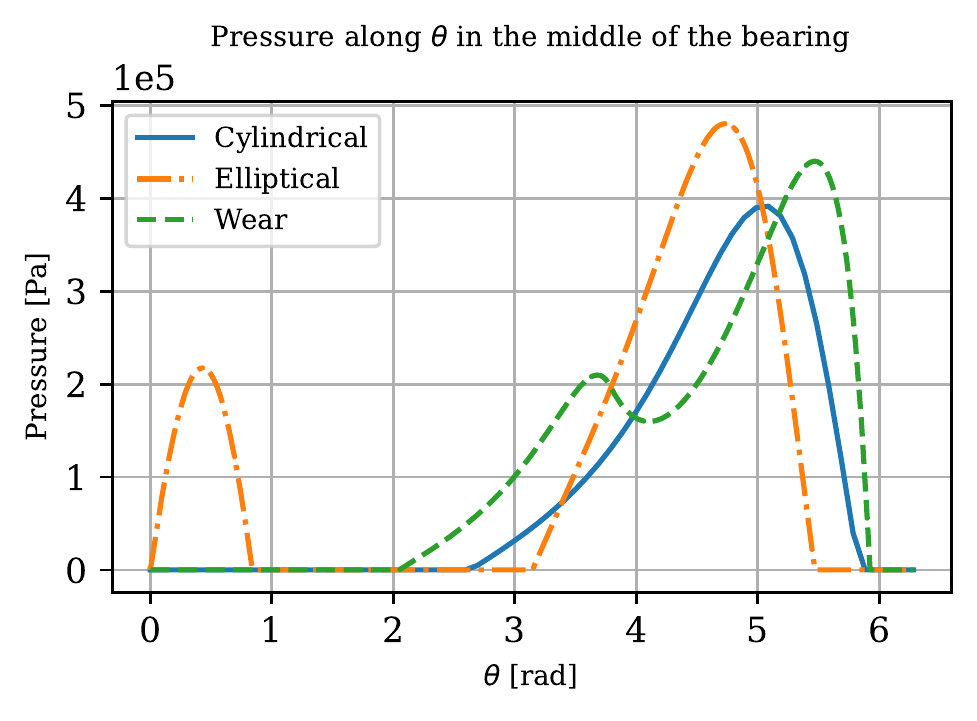}
    \caption{Pressures along $ \theta $ in the middle of the bearing for different geometry configurations, varying the rotation speed.} \label{fig:pressao_cil_elip_wear}
\end{figure}

As already shown, the pressure behavior is changed for each geometry. Fig. \ref{fig:pressao_cil_elip_wear} demonstrates these changes in order to facilitate comparison. It is noticed that, in addition to the new peak, the maximum pressure value is also modified. The cylindrical bearing has the lowest pressure value, while the elliptical reaches the greatest magnitude, even though the configuration with the least eccentricity. It is also possible to notice that the wear causes an increase in the pressure  comparing to the cylindrical bearing, also modifying the place where the pressure reaches its highest value.

\begin{figure} 
    \centering
    \includegraphics{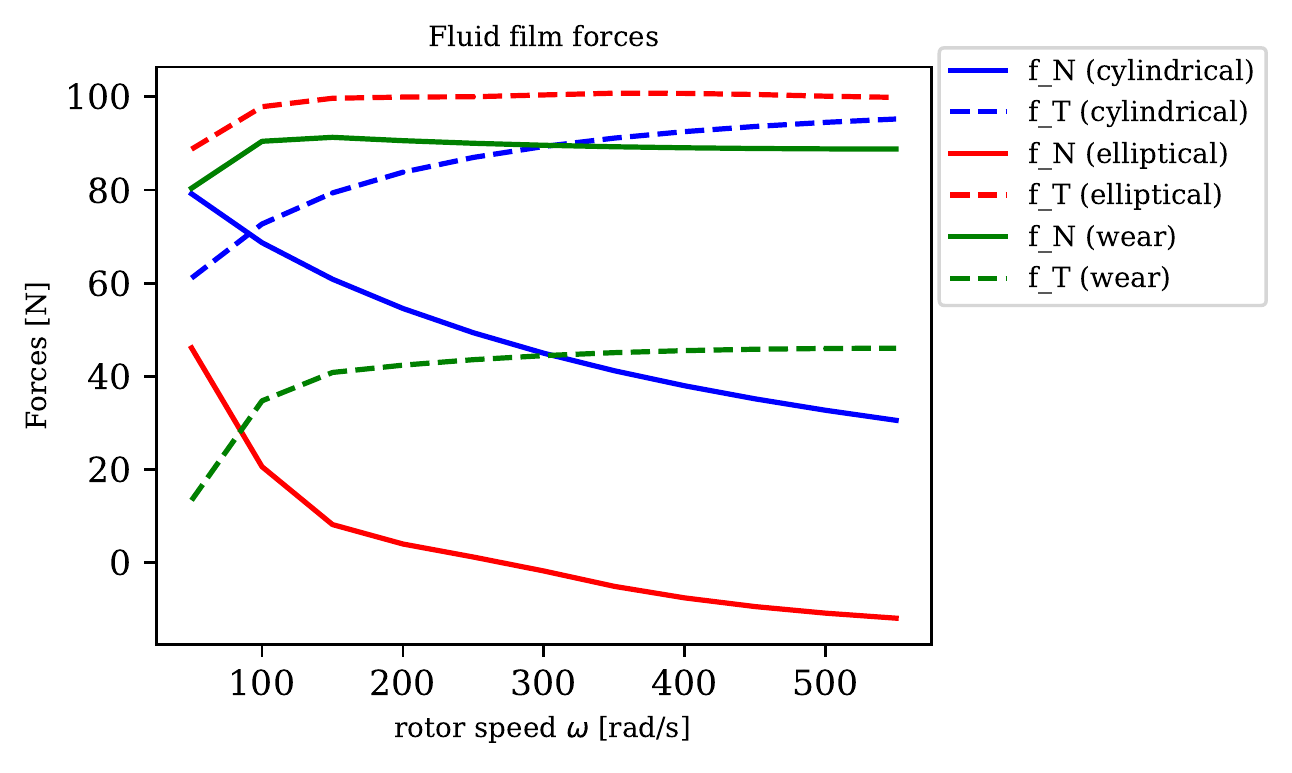}
    \caption{Forces for different bearing configurations, varying the rotation speed. The continuous line represents the radial component and the dashed line represents the tangential component.} \label{fig:forcas_cil_elip_wear}
\end{figure}

The radial and tangential components of the forces of the fluid film (Fig. \ref{fig:forcas_cil_elip_wear}), as well as the pressure fields, are different for the three cases presented. In cylindrical and elliptical bearings, the radial force $ N $ presents a clear decrease when the rotation speed increases, even reaching negative values, for the case of the elliptical bearing. For wear, the decrease of this component is moderate, with values much higher than the tangential force $ T $, which is also growing smoothly. The tangential force $ T $ for the elliptical bearing is little influenced by the rotation speed. From 150 rad/s up, the this force is almost constant for the rotation speed range analyzed. On the other hand, for cylindrical geometry, varying the rotation speed has a higher impact on the forces.

\begin{figure}
    \centering
    \begin{subfigure}[b]{0.49\textwidth}
        \centering
        \includegraphics{ 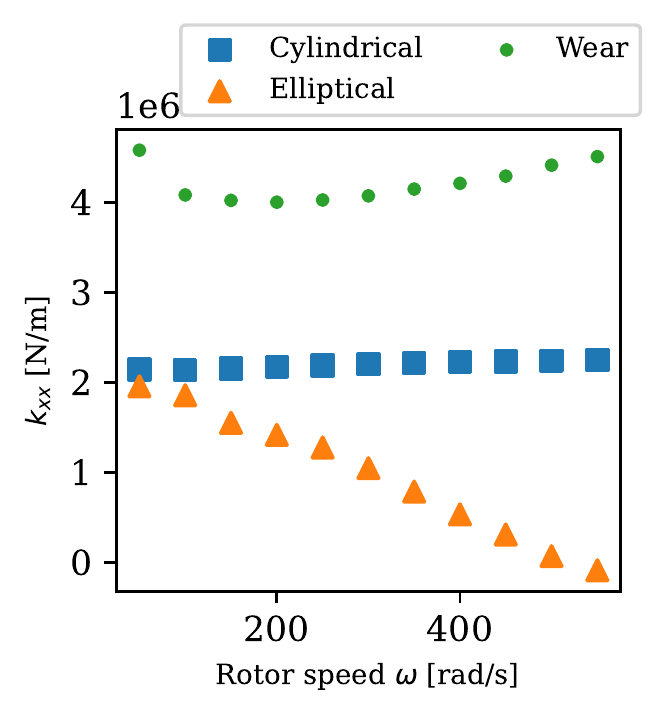}
        \caption{$k_{xx}$}\label{fig:kxx_cil_elip_wear}
    \end{subfigure}
    \begin{subfigure}[b]{0.49\textwidth}
        \centering
        \includegraphics{ 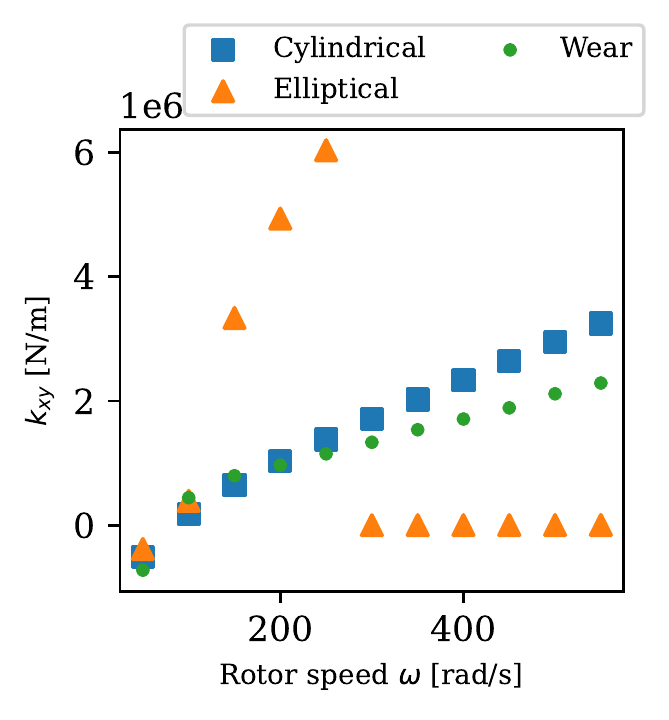}
        \caption{$k_{xy}$}\label{fig:kxy_cil_elip_wear}
    \end{subfigure}\\
    \begin{subfigure}[b]{0.49\textwidth}
        \centering
        \includegraphics{ 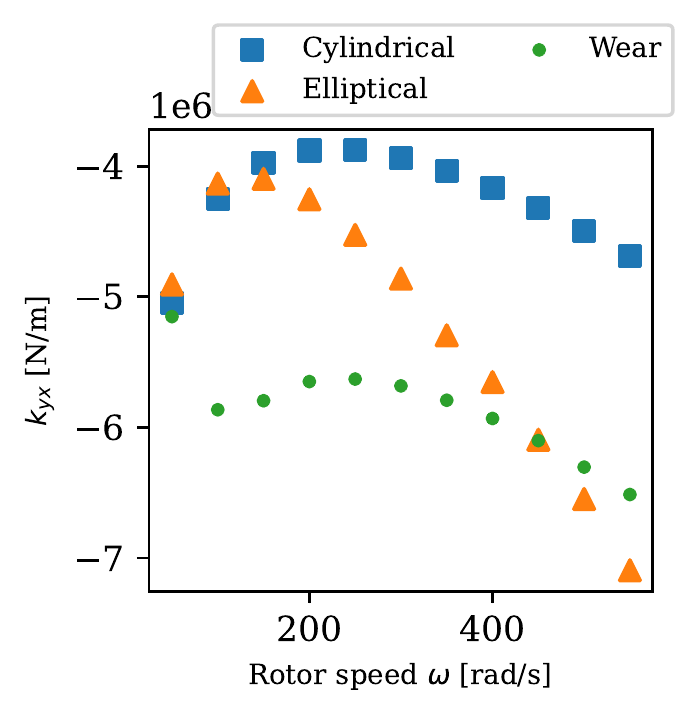}
        \caption{$k_{yx}$}\label{fig:kyx_cil_elip_wear}
    \end{subfigure}
    \begin{subfigure}[b]{0.49\textwidth}
        \centering
        \includegraphics{ 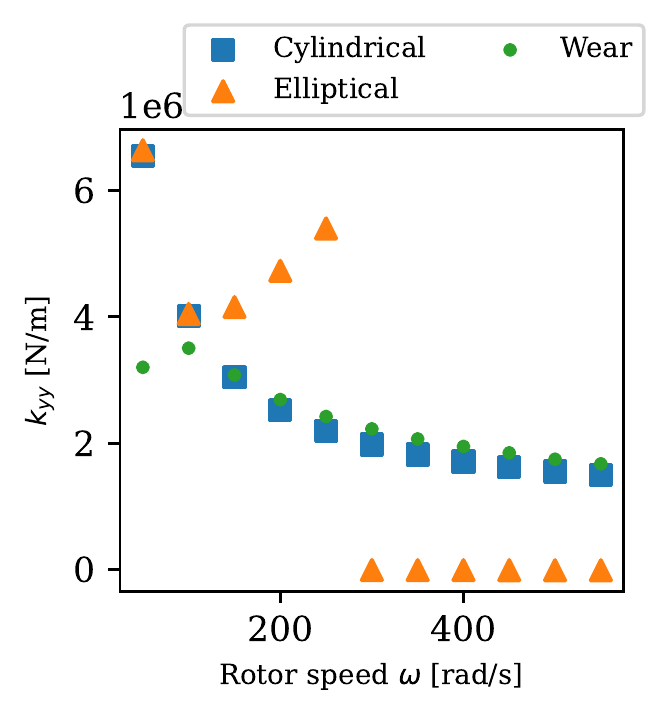}
        \caption{$k_{yy}$}\label{fig:kyy_cil_elip_wear}
    \end{subfigure}
\caption{Stiffness coefficients for different bearing configurations, varying the rotation speed.}\label{fig:rigidez_cil_elip_wear}
\end{figure}

Finally, Figs. \ref{fig:rigidez_cil_elip_wear} (a-d) and \ref{fig:rigidez_cil_elip_wear} (a-d) show the variations of each term of the stiffness and damping coefficients for the different geometries in the rotation speed range adopted. Some characteristics stand out in these results and are commented below.

Regarding the stiffness coefficients, it is noticed that, in the elliptical bearing, the terms $ k_{xy} $ and $ k_{yy} $ have rapid growing up to $ \omega \approx 300 [rad/s] $ and, then there is a drastic reduction in their values, which does not occur in the other two geometries. The term $ k_{xx} $ shows little variation with the rotation speed for the cylindrical bearing. It decreases for the elliptical bearing and develops much higher values for the wear geometry. The terms $ k_{xy} $, $ k_{yx} $ and $ k_{yy} $ show similar growth and decrease their values in cylindrical and worn bearings, with the greatest difference between them in the crossover term $k_{yx}$.

\begin{figure}
    \centering
    \begin{subfigure}[b]{0.49\textwidth}
        \centering
        \includegraphics{ 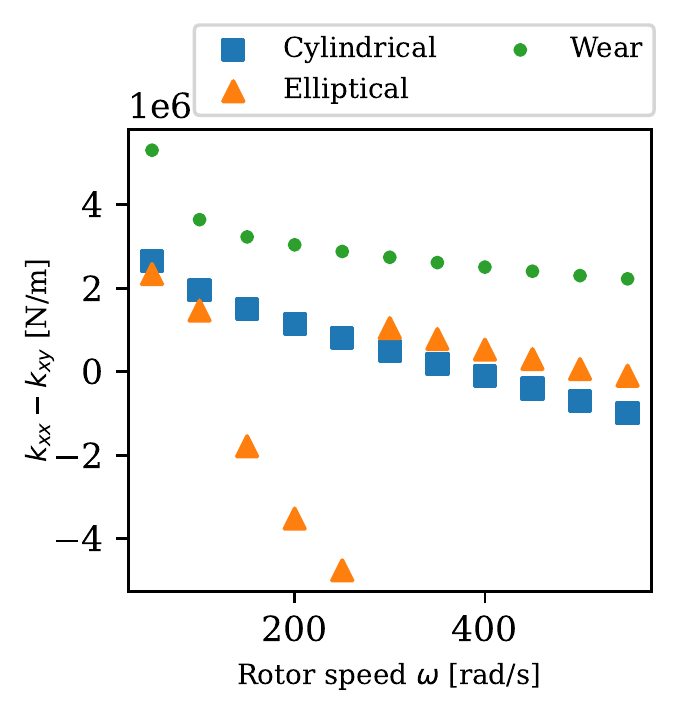}
        \caption{Direction $x$.}
    \end{subfigure}
    \begin{subfigure}[b]{0.49\textwidth}
        \centering
        \includegraphics{ 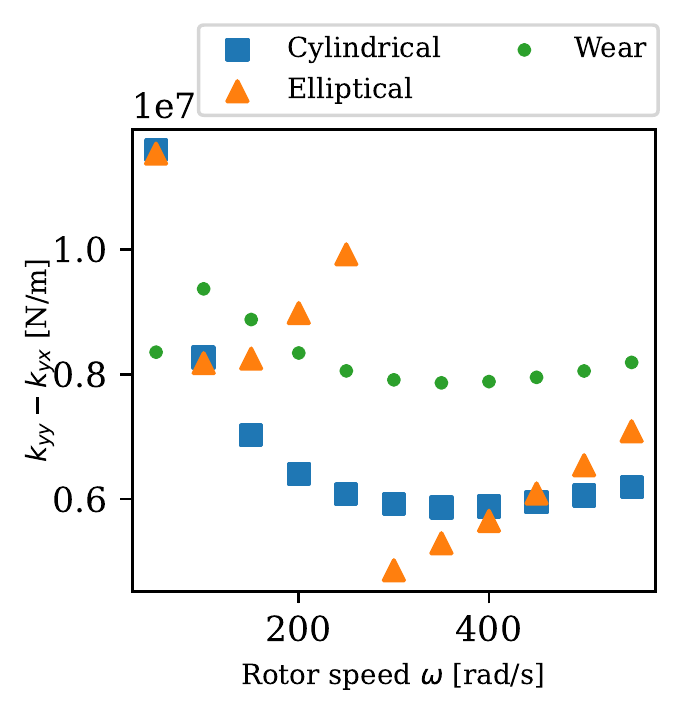}
        \caption{Direction $y$.}
    \end{subfigure}
\caption{Difference between the direct and cross terms of stiffness in each direction.}\label{fig:diferenca-comparacoes}
\end{figure}

In order to analyze the stability conditions in the different bearings, it is necessary to compare the direct and crossed stiffness coefficients for the same direction. Such comparison becomes more understandable in Figs. \ref{fig:diferenca-comparacoes} (a-b), which illustrate the differences $ (k_{xx} - k_{xy}) $ and $ (k_{yy} - k_{yx}) $ in the applied rotation speed range. If these differences are negative, they mean that the cross term has a higher value than the direct term, indicating possibility of rotor instability. It is desired that these values are high and positive. In this way, it is possible to notice that the differences, for the three bearing configurations in the $ x $ direction, tend to decrease. However, the elliptical bearing presents a different behavior from the others, with the cross term greater than the direct one for rotation speeds in the  range $ 100 < \omega <300 \left[rad/s \right] $. The difference in the terms in the $ y $ direction is positive for the entire variation, for the three geometries, also showing much less uniformity in the elliptical bearing.

\begin{figure}
    \centering
    \begin{subfigure}[b]{0.49\textwidth}
        \centering
        \includegraphics{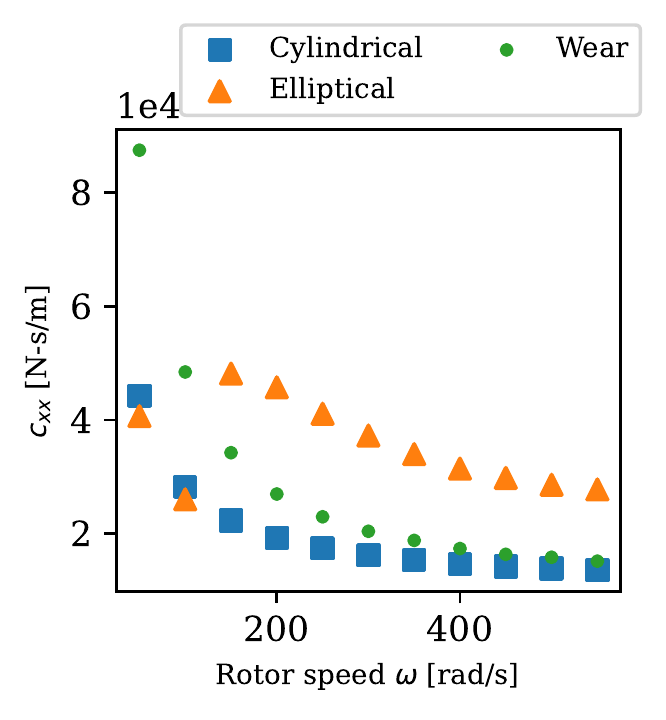}
        \caption{$c_{xx}$}\label{fig:cxx_cil_elip_wear}
    \end{subfigure}
    \begin{subfigure}[b]{0.49\textwidth}
        \centering
        \includegraphics{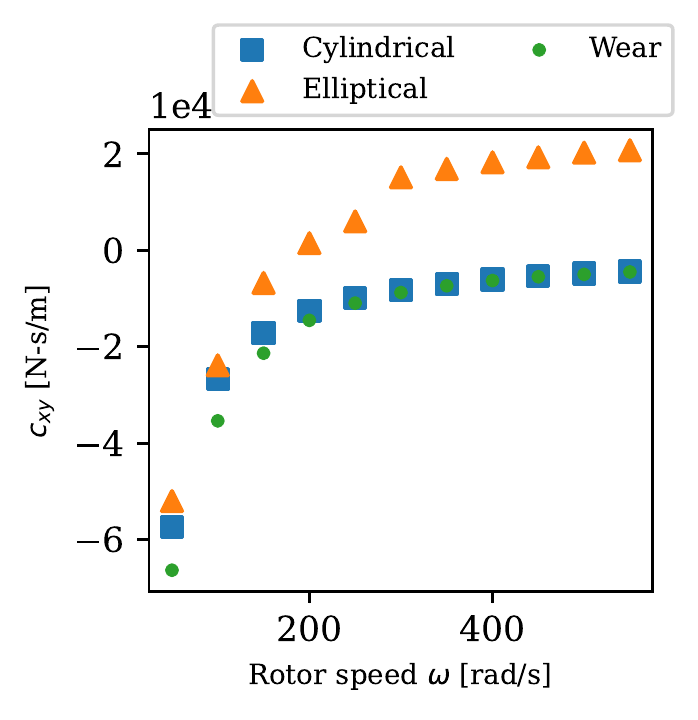}
        \caption{$c_{xy}$}\label{fig:cxy_cil_elip_wear}
    \end{subfigure}\\
    \begin{subfigure}[b]{0.49\textwidth}
        \centering
        \includegraphics{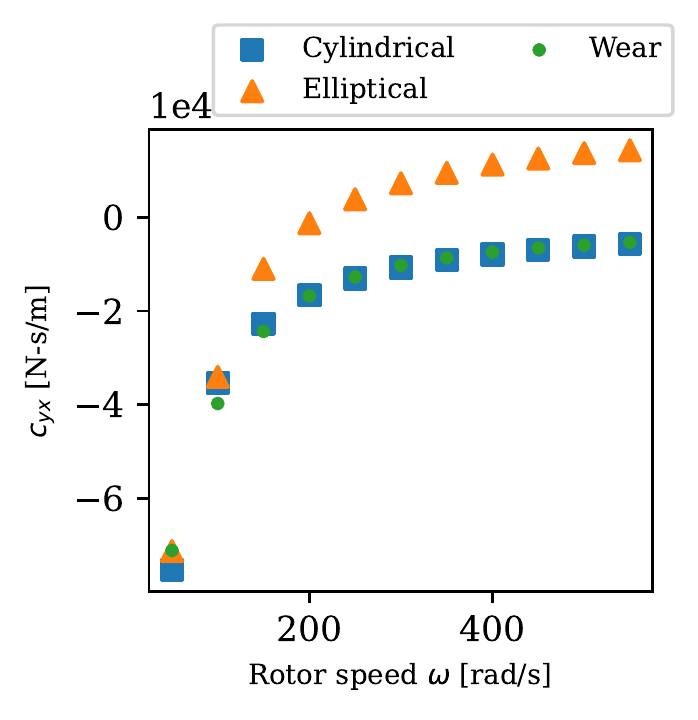}
        \caption{$c_{yx}$}\label{fig:cyx_cil_elip_wear}
    \end{subfigure}
    \begin{subfigure}[b]{0.49\textwidth}
        \centering
        \includegraphics{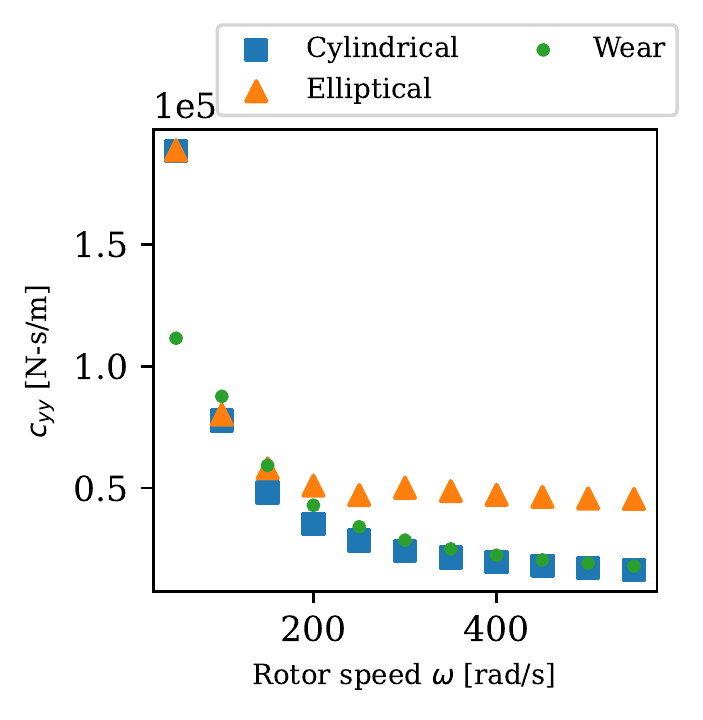}
        \caption{$c_{yy}$}\label{fig:cyy_cil_elip_wear}
    \end{subfigure}
\caption{Damping coefficients for different bearing configurations, varying the speed.}\label{fig:amortecimento_cil_elip_wear}
\end{figure}

Unlike what happens in stiffness, damping is not as sensitive to the presence of faults. All coefficients develop close values for cylindrical and worn bearings, except for the direct terms $ c_{xx} $, which is different for lower speeds, but end up getting closer as the speed increases. The elliptical geometry results in damping coefficients that follow the same growth and decrease patterns as the others, but always with higher values. The cross terms $ c_{xy} $ and $ c_{yx} $ show increasing and negative values for all cases. However, with higher rotation speeds, they become positive only on the elliptical bearing. Direct terms decrease in all cases without reaching negative results.

\section{Conclusions}

Considering the flow of a small thickness of lubricating oil film between two cylinders, this work presents a review of the whole Lubrication Theory, starting from the Navier-Stokes Eq. (\ref{eq:NS_e_continuidade_A}) until reaching an elliptical partial differential Eq. (\ref{eq:reynolds}) that describes the pressure field in the annular space, corresponding to the well-known Reynolds equation normally used in the literature on the subject.

Through a dimensionless analysis, it was possible to promote a mapping of the terms of the equations, ordering them by relevance in the physical phenomenon for this type of situation.  By revisiting this formulation, it was also possible to choose a new and effective parameterization of geometry, which enables a more detailed description of the problem, allowing to adapt the modeling of this work to other contexts. 

Without changing the proposed model, simulations were carried out in different configurations of hydrodynamic bearings. This parameterization also allows the model to be adapted in the future for many other geometries, including changes not only in the tangential direction, but also in the axial direction. In addition, the equations were written in cylindrical coordinates in order to preserve the effects of curvature, which may be interesting for future works.

The pressure field was obtained numerically through the centered finite differences method. This result was verified from its application in simulations for cylindrical hydrodynamic bearings, comparing the responses of this model with those obtained through classic approaches in the literature.

A new methodology was presented, different from the one commonly used in the literature, to compute the stiffness and damping coefficients. This alternative, through a Least Squares solution obtained by the Moore-Penrose pseudo inverse, making the process computationally effective.

Another relevant result is the construction of the algorithm in \textit{Python}, developed by the authors of this research and used in all the simulations presented. Through \textit{Fluid Flow} it is possible to choose geometric and operational parameters according to analyzes of the desired pressure performance, forces, equilibrium position, dynamic coefficients, etc. In addition, the low computational cost of the implemented modeling allows to generate different simulations, which enables its use in machine learning algorithms, a function already explored in \textit{Ross}, which contains a stochastic module to provide this type of analysis. This module also explores uncertainties about the model presented in this work, such as, for example, the impact of the equilibrium position on the dynamic coefficients.

Two other configurations of hydrodynamic bearings were also presented: the elliptical and wear bearing. These new structures, obtained through changes only in the geometric description of the problem, without changing the modeling, portray the gain resulting from the chosen parameterization. In each situation, the impact of these changes on the dynamic response of the system was analyzed and, finally, comparisons were made between the three geometries, exploring the changes in the responses for different rotation speeds.

The results obtained make clear the impact of the variation of the bearing type on the fluid model's responses. The elliptical bearing, introduced as an alternative to correct the instabilities generated in the cylindrical bearing, showed the high stiffness coefficients, which might lead to rotor instability. As for the bearing with wear, it is clear that even small changes compromise the entire response of the model. The pressure field changes dramatically, as do the forces and stiffness. On the other hand, the worn does not seem to influence the system's damping so much.

It is concluded that the description of geometry is fundamental for the success of the modeling. It is worth noting that the parameterization chosen for the description of the problem allows such changes to be made naturally, without the need to modify the model. Furthermore the proposed model may constitute a tool for choosing the best geometry, in order to optimize the operation of rotating machines within their specific contexts.

\section{\emph{Acknowledgements}}

The code used for the simulations presented in this paper is part of ROSS\footnote{\url{https://github.com/ross-rotordynamics/ross}}, an open source library written in \textit{Python} for rotordynamic analysis, by \cite{Ross}.

\appendix


 \bibliographystyle{elsarticle-num} 
 \bibliography{cas-refs}





\end{document}